\documentclass[11pt]{article}

\usepackage[margin=0.75in]{geometry}

\usepackage{changepage}

\usepackage[utf8]{inputenc}

\usepackage[superscript,biblabel]{cite}

\usepackage{nameref,hyperref}

\usepackage{rotating}

\usepackage{setspace} 
\makeatletter
\renewcommand{\@biblabel}[1]{\quad#1.}
\makeatother

\date{}

\usepackage{graphicx}
\usepackage{csvsimple}

\usepackage{booktabs}
\usepackage{algorithm}
\usepackage{algorithmicx}
\usepackage{algpseudocode}
\usepackage{lineno}

\begin{document}
\vspace*{0.35in}

\begin{flushleft}
{\Large
\textbf\newline{Detecting Spatial Patterns of Disease in Large
Collections of Electronic Medical Records Using 
Neighbor-Based Bootstrapping (NB2)}
}
\newline
\\
Maria T Patterson\textsuperscript{1, 2*},
Robert L Grossman\textsuperscript{1, 3, 4, 5**}
\\
\bf{1} Center for Data Intensive Science, University of Chicago, Chicago, IL\\
\bf{2} Department of Astronomy, University of Washington, Seattle, WA*** \\
\bf{3} Computation Institute, University of Chicago, Chicago, IL \\
\bf{4} Section of Computational Biomedicine and Biomedical Data Science, Department of Medicine, University
of Chicago, Chicago, IL\\
\bf{5} Institute for Genomics \& Systems Biology, University of Chicago, Chicago, IL \\

\vskip 0.5in

*Email: maria.t.patterson@gmail.com \\
**Email: robert.grossman@uchicago.edu \\
***MP currently works at the University of Washington \\

\vskip 1in

{\bf keywords:} geospatial variation of disease incidence, geospatial correlation, electronic medical records 
\end{flushleft}

\clearpage
\section*{Abstract}

We introduce a method called neighbor-based bootstrapping (NB2) that can be used to quantify the geospatial variation of a variable.  We applied this method to an analysis of the incidence rates of disease from electronic medical record data (ICD-9 codes) for approximately 100 million individuals in the US over a period of 8 years.  We considered the incidence rate of disease in each county and its geospatially contiguous neighbors and rank ordered diseases in terms of their degree of geospatial variation as quantified by the NB2 method.  

We show that this method yields results in good agreement with established methods for detecting spatial autocorrelation (Moran's I method and kriging).  Moreover, the NB2 method can be tuned to identify both large area and small area geospatial variations.  This method also applies more generally in any parameter space that can be partitioned to consist of regions and their neighbors.

\clearpage

\section*{\uppercase{Introduction}}

As the number of sources and the volume of electronic medical
records (EMR) and electronic health records (EHR) increases, there is a growing ability to aggregate
these data and extract information about population and
public health \cite{Friedman2013, Murdoch2013}.  
Over the past several years, new sources of digital health data, such
as web searchers, social media, mobile phones, and personal health sensors have
increased the number of sources and data volumes even more
\cite{Brownstein2009,Salathe2012,Generous2014}. 
Much of these data can be geocoded with location information so that 
techniques from spatial epidemiology  can be used to explore geospatial variation in disease, health outcomes, 
and population health using disease mapping, disease cluster analysis, and related techniques
\cite{Elliott2004, Rushton2007, Beale2008, Cromley2011}.
With this much geocoded digital health data, there is a need for simple tools and 
algorithms that can be used by researchers across
disciplines for identifying the presence of spatial autocorrelation in
disease incidence data, especially in large datasets \cite{Noble2012}.

An initial starting point for evaluating the presence of patterns in disease or other
geocoded data is determining whether the data are spatially autocorrelated, that is,  
whether the disease rates or values of interest are similar in nearby areas and fall off with distance,
which could indicate the presence of core areas of disease risk
\cite{Becker1998, Tiefelsdorf1998}.

We introduce a Monte Carlo based algorithm that we call Neighbor-based Bootstrapping (NB2)
that can be used to quantify geospatial autocorrelation.  We apply
this algorithm to approximately 100 million geocoded EMR and rank
order 548 diseases as determined by ICD-9 codes from those with the
strongest geospatial autocorrelation to those with the weakest
geospatial autocorrelation.   We compare this method's results to Moran's I
statistic \cite{moran1950} and to kriging \cite[page 44]{ripley2005spatial}, two other techniques that have been used to
quantify geospatial autocorrelation.
The spatial size scale of disease patterns may range widely from small, localized
affected regions to larger affected areas, depending on the nature of the 
underlying factors.  We have developed two versions of the NB2
ranking, one favoring patterns of tight clusters and the other
favoring broader less peaked patterns.

Applying geospatial analysis and visualization techniques to geocoded
health data has long been understood to be important for identifying
risk factors from the physical environment and for providing insights
into the transmission of infectious and vector-borne diseases
\cite{Kitron1998, Hay2000, Moonan2004, Sasaki2008, Kamadjeu2009,
  Nuckols2004, Weis2005}.
For example, spatial analysis of health data can be used to identify and manage risk
associated with proximity to potentially harmful environmental exposures,
such as chemical toxins or air pollutants \cite{Huang1999,Jarup2004,Sasaki2008}.
More generally these techniques are also important for understanding a
broader range of risk factors, including risk factors from the demographic, economic, social,
cultural, regulatory, or  legal environments \cite{Graves2008,
  Dean2010, Harrison2011, Gatrell2014, Luther2003, Geraghty2010,
  Comer2011, Rodriguez2013, Rzhetsky2014}.

\section*{\uppercase{Materials and Methods}}

\subsection*{Data}
The dataset consists of electronic medical records data from the Truven 
Health MarketScan Commercial Claims and Encounters Database,
which includes approved inpatient and outpatient insurance claim information
for a total of approximately 100 million unique and de-identified individuals across the United 
States for the time period from 2003 to 2010.  The records include
1.3 billion diagnostic ICD-9 subdivided codes (12.89 unique codes per person),
geotagged by county FIPS code. Here we restrict to using the approximately 800
non-subdivided ICD-9 codes from 001-799, which excludes injuries, poisonings, 
and accidents.  We refer to ICD-9 codes by the 3 digit integer group.  (For example,
``005: Other bacterial poisoning'' includes ``005.0 Staphylococcal food poisoning''.)

We also restrict our analysis to the 3,109 counties in the continental US and 
to the ICD-9 codes that have data for two-thirds or more of the counties.
This leaves 548 ICD-9 codes.

For each of these 548 codes, we adjust for age and gender 
by using standard populations \cite{klein2001age} as follows.
We determine crude incidence rates for the standard 19 groups of age populations 
for each gender by taking raw counts for each group and dividing by the
population at risk, which in this case we take to be the total number of 
records for each county for each age/gender group converted to 100,000 
person-year units:
\begin{equation}\label{eq:cruderate}
Y_{crude}^{age,gender} = \frac{\mbox{cases of ICD-9}}{\mbox{total cases}} \times \frac{100000 \mbox{ persons}}{8 \mbox{ years}}.
\end{equation}

The age and gender adjusted rate is calculated by multiplying the
crude rate for each group by the appropriate weight using the Census 2010
standard population and summing the products  \cite{Day1992population}:
\begin{equation}\label{eq:adjusrate}
Y_{adjusted} = \Sigma_{age,gender} Y_{crude}^{age,gender} \times \frac{\mbox{group population}}{\mbox{total population}} 
\end{equation}

\subsection*{Neighbor-based bootstrapping (NB2) method}
The NB2 method uses resampling to evaluate in this example
whether or not the incidence rate of a disease can be accurately estimated from 
the incidence rate of the disease in counties that are neighbors. 
The first step in this method is to define regions and neighbors of regions.
Here we define regions as counties and neighboring counties as counties that are geospatially 
contiguous to the county's polygon border, including vertices (Queen style),
though it is important to note that there are many options to consider when defining
neighbor relationships (contiguity, distance, spatial weights) that have varying
effects on results \cite[for example]{Waller2004, Bivand2013}.
In this paper, we focus on geospatially defined neighbors, but
an advantage of this method is that it is applicable without change to neighbors in any space of features,
not just neighbors in 2 or 3 dimensional physical space.

We compute a bootstrapped estimate as follows.
Fix a county Y.  For each ICD-9 code, we sample with replacement
a set of neighboring counties and a set of random counties and compare
the normalized disease incidence (from Eq. \ref{eq:adjusrate}).  

More explicitly, fix a county $Y$ and assume that it has $n^Y$ neighbors.  
We estimate the log incidence rate $Z_{neighbor}$ for county $Y$ as the 
average log incidence rate of a list of $n^Y$ randomly chosen (with replacement) 
neighboring counties.  We also estimate for each county the log incidence rate $Z_{random}$ for county $Y$ as 
the average log incidence rate of $n^Y$  randomly chosen (with replacement)
counties from the full set of all counties.  These counties may or may not be neighbors.

We compare the two estimates (neighbors vs random) to the known log incidence for 
each of the drawn counties in two separate ways.  See
Algorithm~\ref{alg:nb2-ttest} and Algorithm~\ref{alg:nb2-odds}.    

In the first implementation, we take the difference from actual of
the estimates of $Z_{neighbor}$ vs $Z_{random}$.  We then use a paired Student's 
t-test to evaluate whether the neighbor based predictions are a significant improvement 
over the random prediction.  
For ICD-9 codes with significant underlying spatial patterns, we expect that
the $Z_{neighbor}$ estimates will be significantly closer to actual than the $Z_{random}$
estimates.

We repeat this process to obtain 1000 estimates of the neighbor-based
vs random differences, and for each of these compute the paired t-test.   We then take
the median t-test value from these 1000 estimates.
This gives us one t-test statistic value per ICD-9 code, describing how closely related
incidence rates of that ICD-9 are in neighboring counties as compared with a random selection of 
counties.

In the second implementation, we compare the neighbors vs random estimates by counting,
for each pair of bootstraps, the number of samples where the neighbor estimate is closer to
actual than the random estimate.  We then repeat this 1000 times, take the median number,
and using this to calculate the log odds that the neighbor estimate is more accurate than the
random estimate. 

\begin{algorithm}
\caption{Neighbor-based bootstrapping method with paired t-test}
\label{alg:nb2-ttest}
\textbf{INPUT:} Set of records, $\left\{ {counties} \right\}$ with length N; the value of interest 
(log incidence rate $Z$) for each record $Y$; and 
the list of each record's neighbors $\left\{ {neighbors(Y)} \right\}$. \\
\textbf{OUTPUT:} $NB2$ statistic using paired t-test
\begin{algorithmic}
\For {m $\gets$ 1 to M repetitions}

\For {N samples (with replacement) of $Y \in \left\{ {counties} \right\}$}
	\State $Z^{Y} \gets \log$(incidence rate in county Y)
    \State $n^Y \gets$ number of elements in $\left\{ {neighbors(Y)} \right\}$
    \State Choose $n^Y$ counties $\in$ $\left\{ {neighbors(Y)} \right\}$ with replacement, call this $B_{neighbor}$
    \State Choose $n^Y$ counties $\in$ $\left\{ {counties} \right\}$ with replacement, call this $B_{random}$
    \State $Z_{neighbor}^{Y} \gets$ average $Z$ of $B_{neighbor}$
    \State $Z_{random}^{Y} \gets$ average $Z$ of $B_{random}$
\EndFor
	\State $D_{neighbor} \gets$ List of $Z_{neighbor}^Y$ $-$ $Z^Y$for N sampled counties
   	\State $D_{random} \gets$ List of $Z_{random}^Y$ $-$ $Z^Y$ for N sampled counties
    \State Set $t^m$ equal to the paired Student's t-test statistic for $D_{neighbor}$ and $D_{random}$:
    \State $t^m = ( \bar{D}_{neighbor} - \bar{D}_{random})  \sqrt{  \frac{ l ( l-1 ) }  {\sum_{i=1}^{l}  (\hat{D}_{neighbor}^l  - \hat{D}_{random}^l )}}$ 
\EndFor   
	\State $t \gets$ List of $t^m$ for all M repetitions
    \State $NB2$ statistic $=$ median($t$)
\end{algorithmic}
\end{algorithm}

\begin{algorithm}
\caption{Neighbor-based bootstrapping method with log odds}
\label{alg:nb2-odds}
\textbf{INPUT:} Set of records, $\left\{ {counties} \right\}$ with length N; the value of interest 
(log incidence rate $Z$) for each record $Y$; and 
the list of each record's neighbors $\left\{ {neighbors(Y)} \right\}$. \\
\textbf{OUTPUT:} $NB2$ statistic using log odds
\begin{algorithmic}
\For {m $\gets$ 1 to M repetitions}

\For {N samples (with replacement) of $Y \in \left\{ {counties} \right\}$}
	\State $Z^{Y} \gets \log$(incidence rate in county Y)
    \State $n^Y \gets$ number of elements in $\left\{ {neighbors(Y)} \right\}$
    \State Choose $n^Y$ counties $\in$ $\left\{ {neighbors(Y)} \right\}$ with replacement, call this $B_{neighbor}$
    \State Choose $n^Y$ counties $\in$ $\left\{ {counties} \right\}$ with replacement, call this $B_{random}$
    \State $Z_{neighbor}^{Y} \gets$ average $Z$ of $B_{neighbor}$
    \State $Z_{random}^{Y} \gets$ average $Z$ of $B_{random}$
\EndFor
	\State $Z_{neighbor} \gets$ List of $Z_{neighbor}^Y$ for all counties
   	\State $Z_{random} \gets$ List of $Z_{random}^Y$ for all counties
    \State Set $u^m$ equal to the number of samples where the neighbor estimate is closer to actual than the random estimate:
    \State $u^m = $length( abs($Z_{neighbor} - $Y) \textless~ abs($Z_{random} -$Y) == TRUE) )
\EndFor   
	\State $u \gets$ List of $u^m$ for all M repetitions
    \State $NB2$ statistic $= \log \left(\frac{ median(u)} {N - median(u)}\right)$
\end{algorithmic}
\end{algorithm}

\clearpage

\section*{\uppercase{Results}}

\subsection*{Performance}

We first evaluated the impact of varying the number of times $M$ that we resampled.
Running the entire procedure and resampling $M=1000$ times for all 548 diseases 
takes just over 28 hours on a virtual machine
with 8 Xeon cores running at 2.00 GHz with 16 GB of RAM.
This is about  25 minutes per disease using a single core.  

For 100 bootstraps, the run time for 548
diseases on 8 cores takes about 220 minutes, or a little over 3
minutes per disease when using a single core.
Comparing the NB2 statistic values for 1000 vs 100 simulations, the difference on average
is 0.2\% and at maximum 1.6\%.  For 10 bootstraps, the total run time is about 30 minutes,
or about 30 seconds per disease using a single core.  The mean difference between NB2 statistic 
values for 1000 and 10 simulations is 0.2\%, and the maximum
difference is 3.8\%.   The results are summarized in the table below.

\medbreak
\begin{tabular}{|l|l|l|l|}\hline
{\bf \# bootstraps (M) } & {\bf time (min)} & {\bf mean difference} & {\bf max difference}  \\ \hline
1000 & 25 & NA & NA \\ \hline
100 &   3   &  0.2\%   & 1.6\% \\ \hline
10 &     0.5 & 0.2\%   & 3.8\% \\  \hline     
\end{tabular}
\medbreak

In the analysis that follows, we are primarily focused on the rank ordering of the ICD-9
codes according to these two implementations of the NB2 method.  
There is no significant difference in the rank orderings between 1000 and 100 or 10 repeated bootstraps.

\subsection*{Comparison with Moran's I statistic}

We compare the neighbor based bootstrapping results to the
global Moran's \textit{I} statistic for detecting spatial autocorrelation, 
which is based on the sum over weights between units multiplied by the
mean-adjusted outcome of interest divided by the squared mean difference of each point. 
Moran's \textit{I} is defined as \cite{moran1950}:
\begin{equation}\label{eq:moran}
I = \frac{n}{\Sigma_i\Sigma_j w_{ij}} \frac{\Sigma_i\Sigma_j w_{ij} (y_i - \bar{y}) (y_j - \bar{y}) } {\Sigma_i (y_i - \bar{y})^2}
\end{equation}
where n is the total number of spatial polygons (counties), $y_i$ is the
value of interest of the $i$th polygon, $\bar{y}$ is the global mean, 
and $w_{ij}$ is the spatial weight of the link between polygon $i$ and $j$. 

\begin{figure}[!ht] 
\centering
\includegraphics[width=0.44\textwidth]{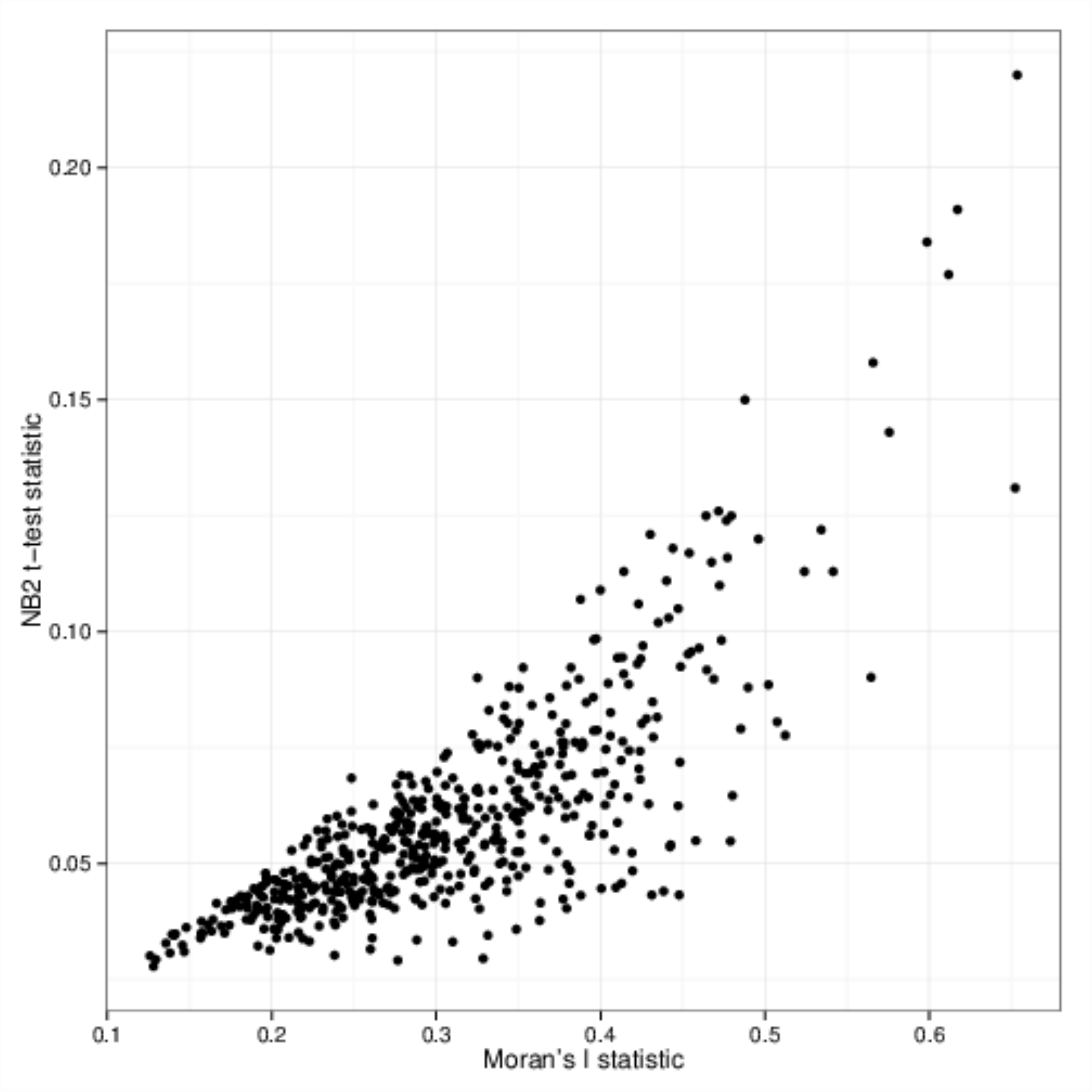}
\includegraphics[width=0.44\textwidth]{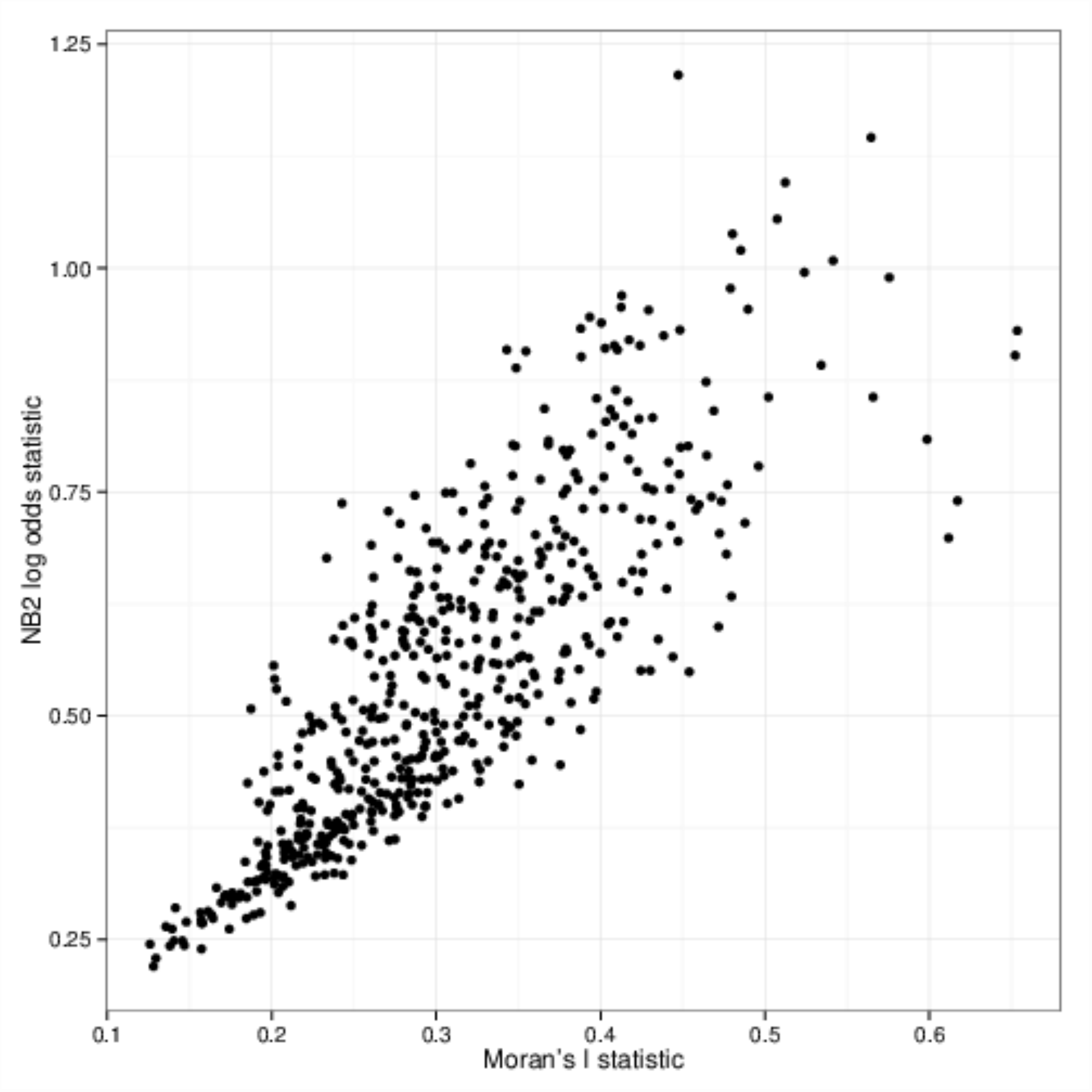}
\caption{{\bf Comparison of 
this neighbor-based bootstrapping (NB2) method
with Moran's \textit{I} statistic for detecting spatial autocorrelation.}
We show the NB2 experiment's ability to detect spatial correlation (y-axis)
measured by the paired Student's t-test estimate (left) 
and the log odds estimate (right) for the neighbor 
county predictions vs random county predictions
plotted against the Moran's \textit{I} statistic estimate (x-axis).
}
\label{Fig1}
\end{figure}

Moran's \textit{I} ranges from $-$1 (perfect dispersion, as in a black and white
checkerboard pattern) to 1 (black squares on one side, white on the other).  
A random distribution would have \textit{I} close to 0.
We compare the values of Moran's \textit{I} for the set of log incidence rates
across counties for each ICD-9 code to both the implementation of the NB2 method using
the paired Student's t-test evaluation
and the implementation using the log odds evaluation.  
If the geospatial variation that the NB2 method detects is similar to
Moran's \textit{I}, then the NB2 statistic values should increase as Moran's 
\textit{I} goes to 1.  
In Figure \ref{Fig1}, we show the NB2 t-test statistic estimate (left) and the log odds estimate (right)
plotted against Moran's I statistic for all ICD-9 codes tested. Generally,
the NB2 statistic values increase as Moran's I statistic increases, though there
is noticeable scatter.

We rank ordered the ICD-9 codes using the two NB2 method implementations and the Moran's I statistic
to produce three ordered lists of ICD-9 codes from the strongest spatial correlation 
(largest Moran's I statistic,  largest NB2 t-test, largest NB2 log odds test) to the weakest.  
Tables \ref{tab:NB2Rank} - \ref{tab:MoranRank} contain the top 25 ICD-9 codes
for both NB2 implementations and Moran's I.  Here we will compare the properties
of the spatial distributions for ICD-9 codes ranked highly by the two NB2 procedures
with those ranked highly by Moran's I statistic.


\begin{sidewaystable}
 \caption{
 {\bf Top 25 ICD-9 codes as ranked by NB2 (t-test)}}
 	\begin{filecontents*}{Table1.txt}
rank.mc,rank.moran,icd9,ttest.median,moran.I.estimate,cat1,cat3,model2.range,sill,rank.mc.odds
1,1.5,402,0.22,0.65,Hypertensive heart disease,Diseases of the Circulatory System,1890,2.32,19
2,3,131,0.191,0.62,Trichomoniasis,Infectious and Parasitic Diseases,610,1.33,85
3,5,635,0.184,0.6,Legally induced abortion,Complications Of Pregnancy Childbirth And The Puerperium,330,1.13,48
4,4,088,0.177,0.61,Other arthropod-borne diseases,Infectious and Parasitic Diseases,580,1.41,110
5,7,115,0.158,0.57,Histoplasmosis,Infectious and Parasitic Diseases,830,1.55,34.5
6,17,219,0.15,0.49,Other benign neoplasm of uterus,Neoplasms,90,0.95,101
7,6,413,0.143,0.58,Angina pectoris,Diseases of the Circulatory System,790,0.7,9
8,1.5,739,0.131,0.65,Nonallopathic lesions not elsewhere classified,Diseases of The Musculoskeletal System And Connective Tissue,490,0.59,28
9,26,602,0.126,0.47,Other disorders of prostate,Diseases of the Genitourinary System,120,0.81,202
10.5,21,099,0.125,0.48,Other venereal diseases,Infectious and Parasitic Diseases,280,0.66,167
10.5,31,520,0.125,0.46,Disorders of tooth development and eruption,Diseases of the Digestive System,130,0.45,32
12,21,009,0.124,0.48,Ill-defined intestinal infections,Infectious and Parasitic Diseases,190,0.64,127.5
13,10,411,0.122,0.53,Other acute and subacute forms of ischemic heart disease,Diseases of the Circulatory System,1020,0.6,30
14,51.5,763,0.121,0.43,Fetus or newborn affected by other complications of labor and delivery,Certain Conditions Originating In The Perinatal Period,60,0.78,252.5
15,14.5,281,0.12,0.5,Other deficiency anemias,Diseases of the Blood and Blood-forming Organs,1090,0.69,63
16,44,042,0.118,0.44,Human immunodeficiency virus (HIV) infection,Infectious and Parasitic Diseases,410,0.72,238
17,37,662,0.117,0.45,Long labor,Complications Of Pregnancy Childbirth And The Puerperium,100,1.01,254.5
18,21,514,0.116,0.48,Pulmonary congestion and hypostasis,Diseases of the Respiratory System,480,0.58,71
19,26,268,0.115,0.47,Vitamin D deficiency,Endocrine Nutritional And Metabolic Diseases And Immunity Disorders,310,0.54,82
21,74.5,066,0.113,0.41,Other arthropod-borne viral diseases,Infectious and Parasitic Diseases,370,1.34,196
21,9,424,0.113,0.54,Other diseases of endocardium,Diseases of the Circulatory System,570,0.4,7
21,11,487,0.113,0.52,Influenza,Diseases of the Respiratory System,830,0.46,8
23,44,135,0.111,0.44,Sarcoidosis,Infectious and Parasitic Diseases,540,0.49,161
24,26,259,0.11,0.47,Other endocrine disorders,Endocrine Nutritional And Metabolic Diseases And Immunity Disorders,200,0.49,107
25,88,476,0.109,0.4,Chronic laryngitis and laryngotracheitis,Diseases of the Respiratory System,50,0.62,231.5
\end{filecontents*}
\csvreader[tabular=|c |c | c  | l | c | r | l|,
     	table head = \hline \textbf{NB2 (t-test)} & \textbf{NB2 (odds)} &\textbf{Moran} &  \textbf{ICD-9 diagnosis name} & \textbf{ICD-9} &\textbf{Range} &\textbf{Sill}\\\hline,
         table foot = \hline]
     {Table1.txt}{1=\rankmc, 2=\rankmoran, 3=\icd, 4=\mc, 5=\moran, 6=\dis, 7=\discat, 8=\range,9=\sill,10=\rankodds}
     { \rankmc & \rankodds & \rankmoran &  \dis & \icd & \range & \sill}
 \label{tab:NB2Rank}
 \end{sidewaystable}

 \begin{sidewaystable}
 \caption{
 {\bf Top 25 ICD-9 codes as ranked by NB2 (log odds)}}
 	\begin{filecontents*}{Table2.txt}
rank.mc.odds,rank.mc,rank.moran,icd9,ttest.median,moran.I.estimate,cat1,cat3,model2.range,sill,odds.ratio
1,171,37,401,0.0625,0.45,Essential hypertension,Diseases of the Circulatory System,530,0.07,3.3727
2,47,8,477,0.0902,0.56,Allergic rhinitis,Diseases of the Respiratory System,630,0.19,3.1453
3,81,12.5,682,0.0777,0.51,Other cellulitis and abscess,Diseases of The Skin And Subcutaneous Tissue,530,0.14,2.991
4,70,12.5,627,0.0806,0.51,Menopausal and postmenopausal disorders,Diseases of the Genitourinary System,530,0.15,2.8717
5,149,21,530,0.0647,0.48,Diseases of esophagus,Diseases of the Digestive System,530,0.11,2.8241
6,75,17,616,0.0791,0.49,Inflammatory disease of cervix vagina and vulva,Diseases of the Genitourinary System,630,0.22,2.7731
7,21,9,424,0.113,0.54,Other diseases of endocardium,Diseases of the Circulatory System,570,0.4,2.7413
8,21,11,487,0.113,0.52,Influenza,Diseases of the Respiratory System,830,0.46,2.7056
9,7,6,413,0.143,0.58,Angina pectoris,Diseases of the Circulatory System,790,0.7,2.6902
10,261.5,21,599,0.0549,0.48,Other disorders of urethra and urinary tract,Diseases of the Genitourinary System,530,0.08,2.6576
11,383,74.5,272,0.0458,0.41,Disorders of lipoid metabolism,Endocrine Nutritional And Metabolic Diseases And Immunity Disorders,530,0.06,2.6363
12,108,74.5,414,0.0723,0.41,Other forms of chronic ischemic heart disease,Diseases of the Circulatory System,530,0.15,2.6025
13,56,17,535,0.088,0.49,Gastritis and duodenitis,Diseases of the Digestive System,720,0.3,2.5963
14,165.5,51.5,692,0.0629,0.43,Contact dermatitis and other eczema,Diseases of The Skin And Subcutaneous Tissue,530,0.12,2.5942
15,245.5,101,785,0.0561,0.39,Symptoms involving cardiovascular system,Symptoms Signs and Ill-defined Conditions,530,0.09,2.5736
16,401.5,88,789,0.0447,0.4,Other symptoms involving abdomen and pelvis,Symptoms Signs and Ill-defined Conditions,530,0.06,2.5572
17,431,101,786,0.0432,0.39,Symptoms involving respiratory system and other chest symptoms,Symptoms Signs and Ill-defined Conditions,530,0.05,2.541
18,110,37,794,0.0719,0.45,Nonspecific abnormal results of function studies,Symptoms Signs and Ill-defined Conditions,530,0.15,2.537
19,1,1.5,402,0.22,0.65,Hypertensive heart disease,Diseases of the Circulatory System,1890,2.32,2.535
20,413,44,780,0.0441,0.44,General symptoms,Symptoms Signs and Ill-defined Conditions,530,0.05,2.521
21,101,61.5,285,0.0744,0.42,Other and unspecified anemias,Diseases of the Blood and Blood-forming Organs,630,0.19,2.509
22.5,289,74.5,250,0.053,0.41,Diabetes mellitus,Endocrine Nutritional And Metabolic Diseases And Immunity Disorders,530,0.09,2.4933
22.5,102,61.5,592,0.0743,0.42,Calculus of kidney and ureter,Diseases of the Genitourinary System,630,0.17,2.4933
24,168.5,88,110,0.0627,0.4,Dermatophytosis,Infectious and Parasitic Diseases,530,0.14,2.4854
25.5,210,74.5,564,0.0589,0.41,Functional digestive disorders not elsewhere classified,Diseases of the Digestive System,530,0.1,2.4815
\end{filecontents*}
\csvreader[tabular=|c |c | c | l | c | r | l|,
     	table head = \hline \textbf{NB2 (odds)} & \textbf{NB2 (t-test)} & \textbf{Moran} &  \textbf{ICD-9 diagnosis name} & \textbf{ICD-9} &\textbf{Range} &\textbf{Sill}\\\hline,
         table foot = \hline]
     {Table2.txt}{1=\rankodds, 2=\rankmc, 3=\rankmoran, 4=\icd, 5=\mc, 6=\moran, 7=\dis, 8=\discat, 9=\range,10=\sill}
     { \rankodds & \rankmc & \rankmoran &  \dis & \icd & \range & \sill}
 \label{tab:NB2Rankodds}
 \end{sidewaystable}

 \begin{sidewaystable}
 \caption{
 {\bf Top 25 ICD-9 codes as ranked by Moran's I}}
 	\begin{filecontents*}{Table3.txt}
rank.moran,rank.mc,icd9,moran.I.estimate,ttest.median,cat1,cat3,model2.range,sill,rank.mc.odds
1.5,1,402,0.65,0.22,Hypertensive heart disease,Diseases of the Circulatory System,1890,2.32,19
1.5,8,739,0.65,0.131,Nonallopathic lesions not elsewhere classified,Diseases of The Musculoskeletal System And Connective Tissue,490,0.59,28
3,2,131,0.62,0.191,Trichomoniasis,Infectious and Parasitic Diseases,610,1.33,85
4,4,088,0.61,0.177,Other arthropod-borne diseases,Infectious and Parasitic Diseases,580,1.41,110
5,3,635,0.6,0.184,Legally induced abortion,Complications Of Pregnancy Childbirth And The Puerperium,330,1.13,48
6,7,413,0.58,0.143,Angina pectoris,Diseases of the Circulatory System,790,0.7,9
7,5,115,0.57,0.158,Histoplasmosis,Infectious and Parasitic Diseases,830,1.55,34.5
8,47,477,0.56,0.0902,Allergic rhinitis,Diseases of the Respiratory System,630,0.19,2
9,21,424,0.54,0.113,Other diseases of endocardium,Diseases of the Circulatory System,570,0.4,7
10,13,411,0.53,0.122,Other acute and subacute forms of ischemic heart disease,Diseases of the Circulatory System,1020,0.6,30
11,21,487,0.52,0.113,Influenza,Diseases of the Respiratory System,830,0.46,8
12.5,70,627,0.51,0.0806,Menopausal and postmenopausal disorders,Diseases of the Genitourinary System,530,0.15,4
12.5,81,682,0.51,0.0777,Other cellulitis and abscess,Diseases of The Skin And Subcutaneous Tissue,530,0.14,3
14.5,53,238,0.5,0.0886,Neoplasm of uncertain behavior of other and unspecified sites and tissues,Neoplasms,220,0.3,34.5
14.5,15,281,0.5,0.12,Other deficiency anemias,Diseases of the Blood and Blood-forming Organs,1090,0.69,63
17,6,219,0.49,0.15,Other benign neoplasm of uterus,Neoplasms,90,0.95,101
17,56,535,0.49,0.088,Gastritis and duodenitis,Diseases of the Digestive System,720,0.3,13
17,75,616,0.49,0.0791,Inflammatory disease of cervix vagina and vulva,Diseases of the Genitourinary System,630,0.22,6
21,12,009,0.48,0.124,Ill-defined intestinal infections,Infectious and Parasitic Diseases,190,0.64,127.5
21,10.5,099,0.48,0.125,Other venereal diseases,Infectious and Parasitic Diseases,280,0.66,167
21,18,514,0.48,0.116,Pulmonary congestion and hypostasis,Diseases of the Respiratory System,480,0.58,71
21,149,530,0.48,0.0647,Diseases of esophagus,Diseases of the Digestive System,530,0.11,5
21,261.5,599,0.48,0.0549,Other disorders of urethra and urinary tract,Diseases of the Genitourinary System,530,0.08,10
26,24,259,0.47,0.11,Other endocrine disorders,Endocrine Nutritional And Metabolic Diseases And Immunity Disorders,200,0.49,107
26,19,268,0.47,0.115,Vitamin D deficiency,Endocrine Nutritional And Metabolic Diseases And Immunity Disorders,310,0.54,82
\end{filecontents*}
\csvreader[tabular=|c |c | c | l | c | r | l|,
     	table head = \hline \textbf{Moran} & \textbf{NB2 (t-test)}  & \textbf{NB2 (odds)}  & \textbf{ICD-9 diagnosis name} & \textbf{ICD-9} & \textbf{Range} & \textbf{Sill}\\\hline,
         table foot = \hline]
     {Table3.txt}{1=\rankmoran, 2=\rankmc, 3=\icd, 4=\mc, 5=\moran, 6=\dis, 7=\discat, 8=\range,9=\sill, 10=\rankodds}
     { \rankmoran & \rankmc & \rankodds &  \dis & \icd & \range & \sill}
 \label{tab:MoranRank}
 \end{sidewaystable}

\subsection*{Scale of spatial influence}
We applied a geostatistical ordinary kriging procedure 
using the R package \verb+automap+ to fit semivariograms models describing
the spatial variation across the continental US for the incidence rates
of each of the ICD-9 diagnostic codes.
The semivariograms show the mean semivariance of values in binned separation distances 
between all pairs of spatial points.  Here we use as input values the log incidence of the 
given ICD-9 in a county and approximate the spatial location of the observation as the county population 
centroid given by the US 2010 Census.  

The semivariogram describes the distance within which the incidence rate is spatially
autocorrelated.
At separation distances where the semivariance is low, points have similar incidence rates.  
To quantify the size of spatial variation, we fit exponential semivariogram models to the data.  
The semivariogram model range describes the distance at which the model flattens to a constant
semivariance.  The semivariogram model sill describes the semivariance value at the range.

\begin{figure}[!ht] 
\centering
%

\includegraphics[width=0.475\textwidth]{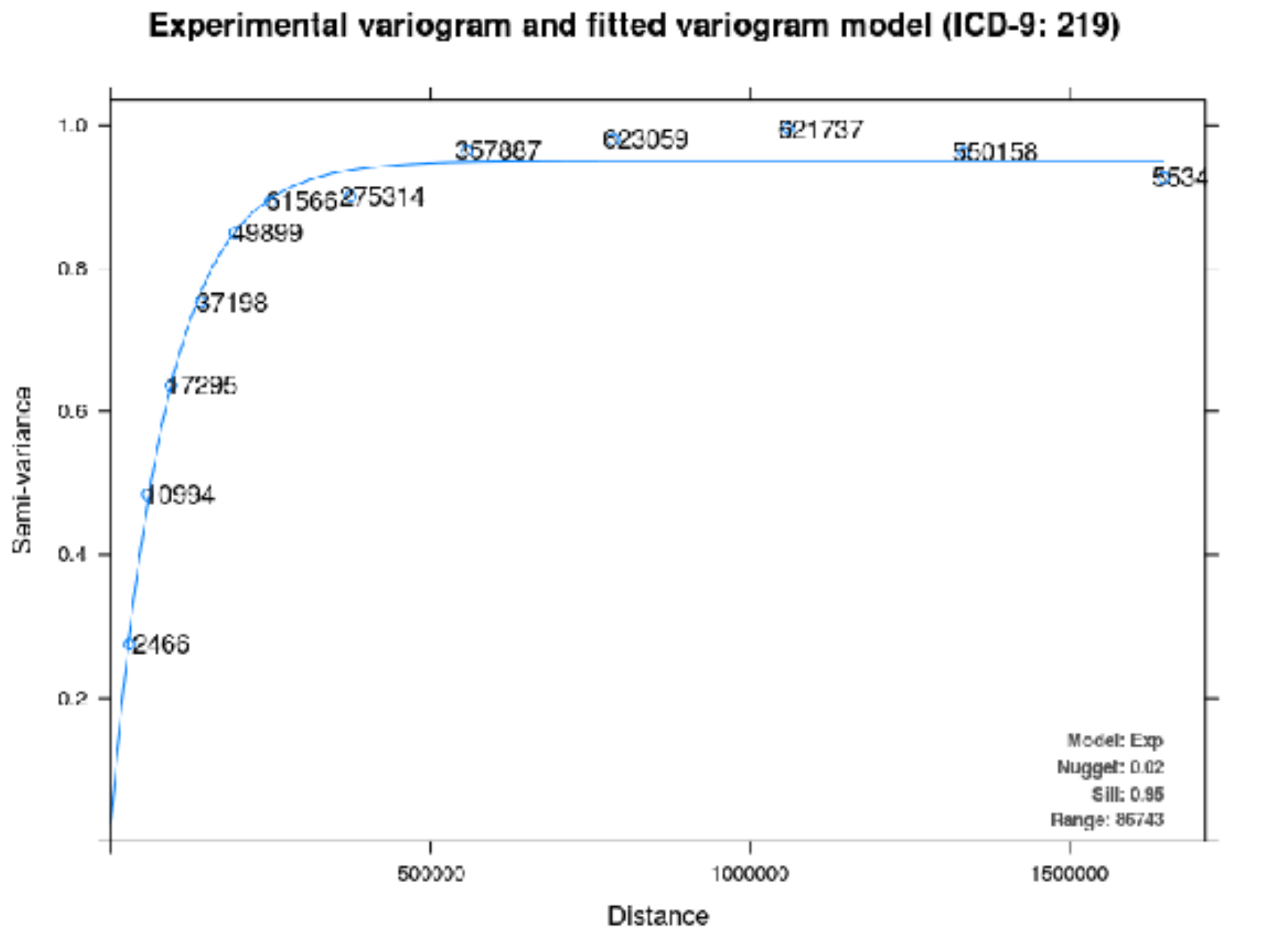}
\includegraphics[width=0.475\textwidth]{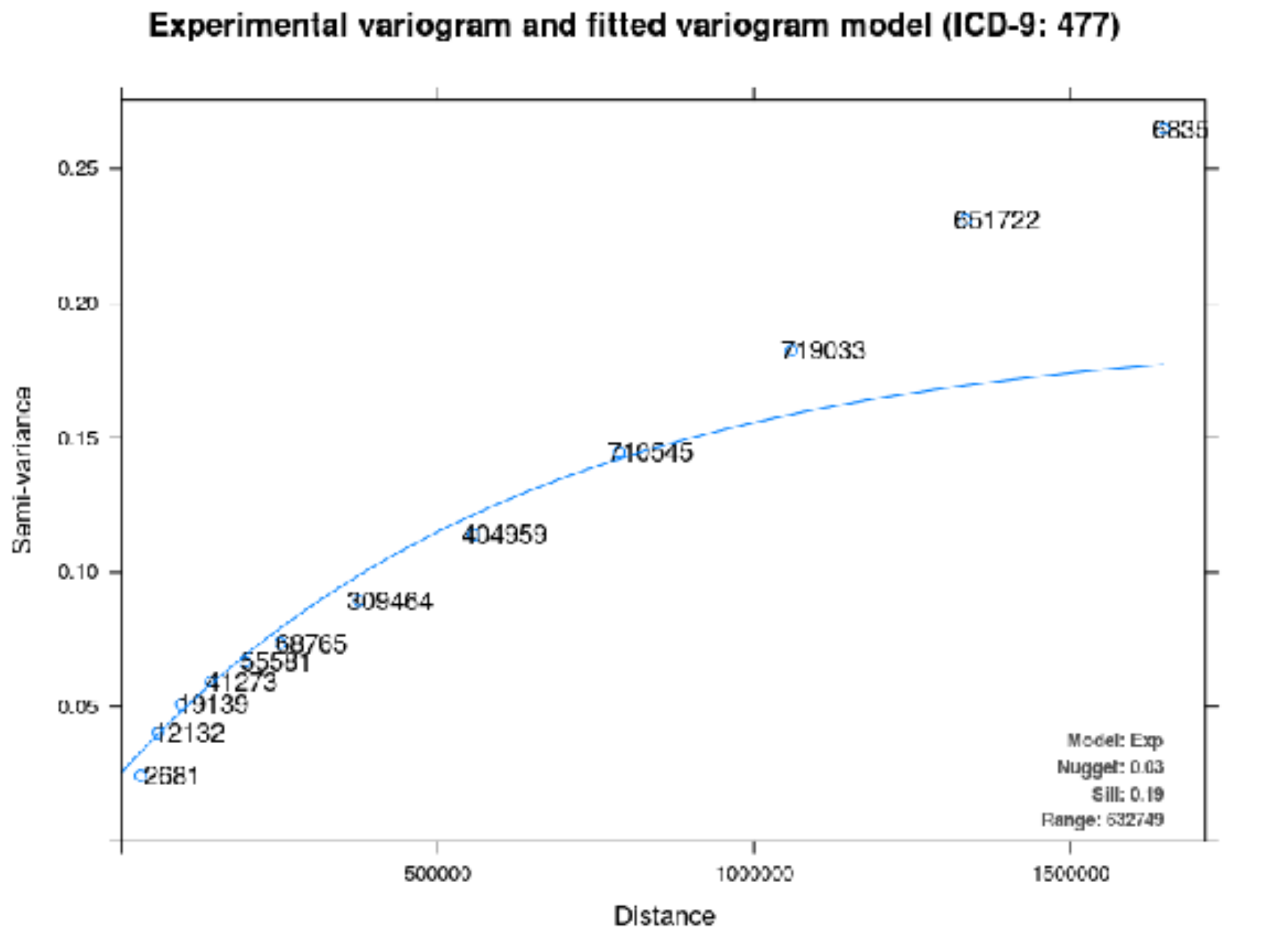}
\includegraphics[width=0.475\textwidth]{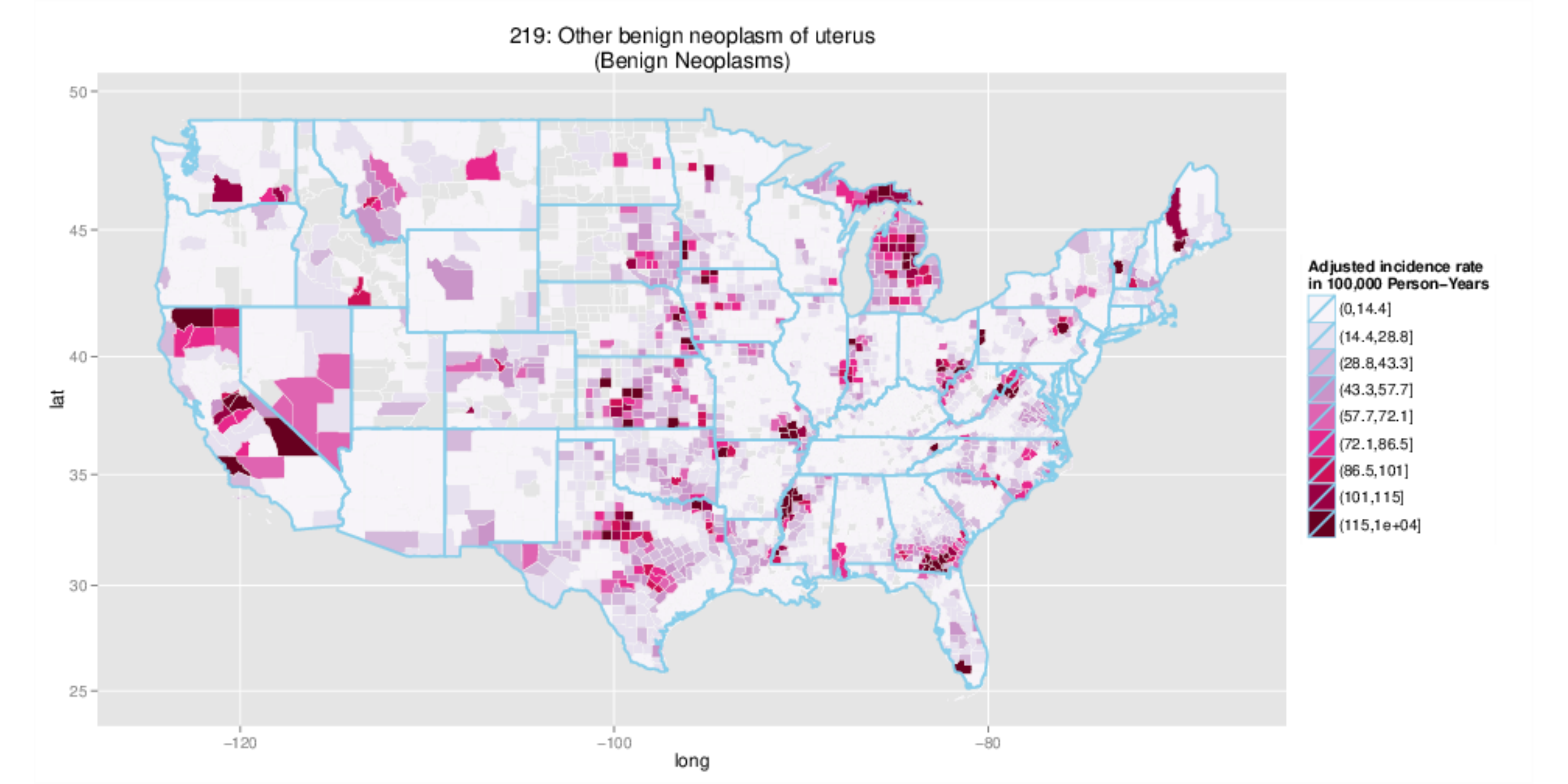}
\includegraphics[width=0.475\textwidth]{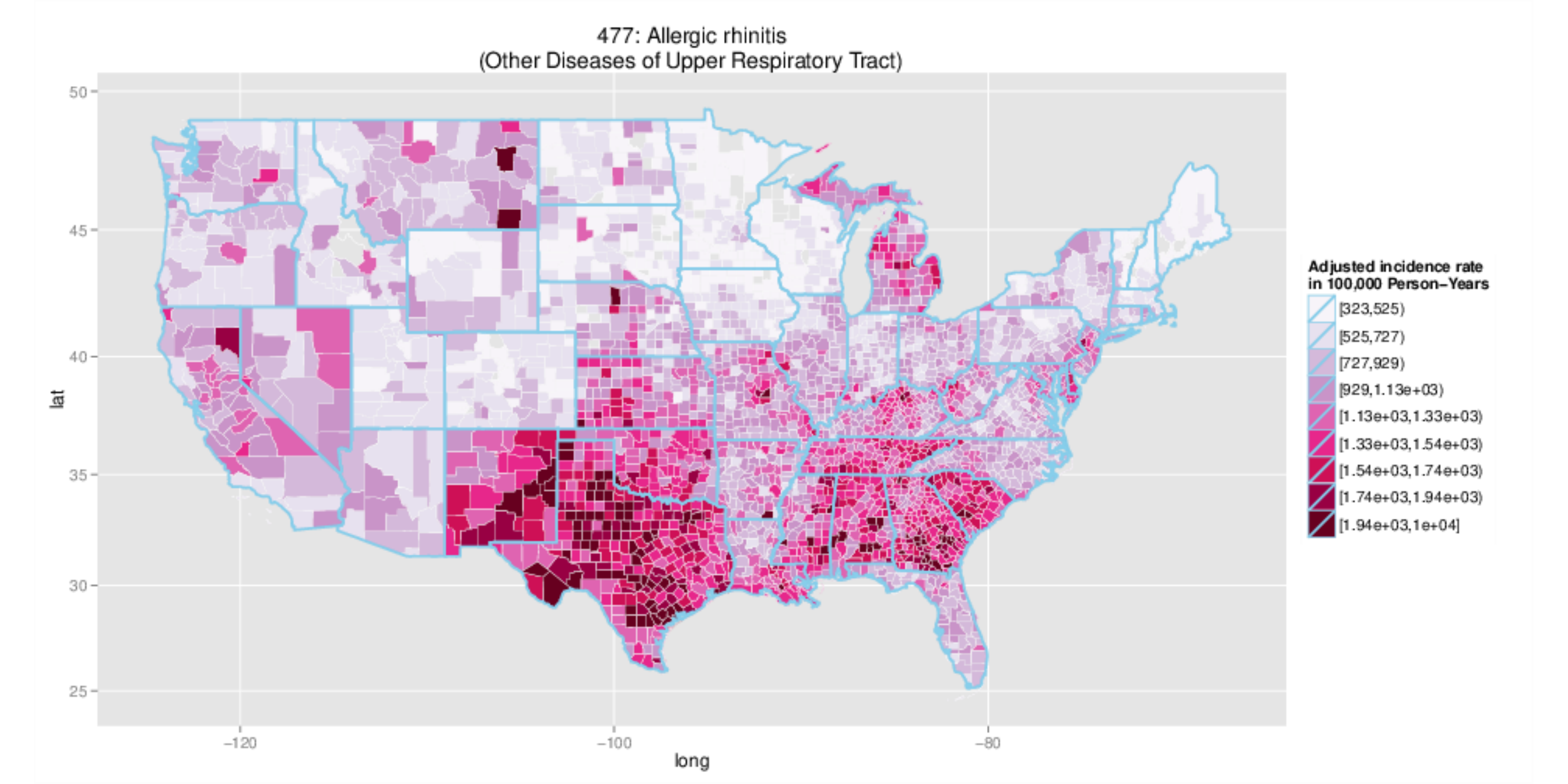}
\caption{{\bf Example semivariograms and incidence rate maps for two ICD-9 codes. }
Top: Semivariograms for ICD-9 219, Other benign neoplasms of uterus, (left) and 
477, Allergic rhinitis, (right).  Bottom: Corresponding incidence rate maps.
}
\label{Fig2}
\end{figure}

In Figure \ref{Fig2}, we show two sample semivariograms, one for an ICD-9 code ranked 
highly by both the NB2 t-test implementation and Moran's I (219: Other benign neoplasms of uterus) 
but relatively low by the NB2 log odds implementations
and one for an ICD-9 code ranked highly by both the NB2 log odds implementation and Moran's I 
but relatively low by the NB2 t-test implementation (477: Allergic rhinitis).
The semivariogram model for 219 has a steep rise that quickly flattens (shorter range), 
and the semivariogram model for 477 continues to rise at large distance.
The incidence rate maps in the bottom of Figure \ref{Fig2} correspondingly show smaller, high peaked 
cluster patterns of spatial variation for 219 (top left) and a larger scale gradation for 477
(top right).

\begin{figure}[!ht] 
\centering

%

\includegraphics[width=0.45\textwidth]{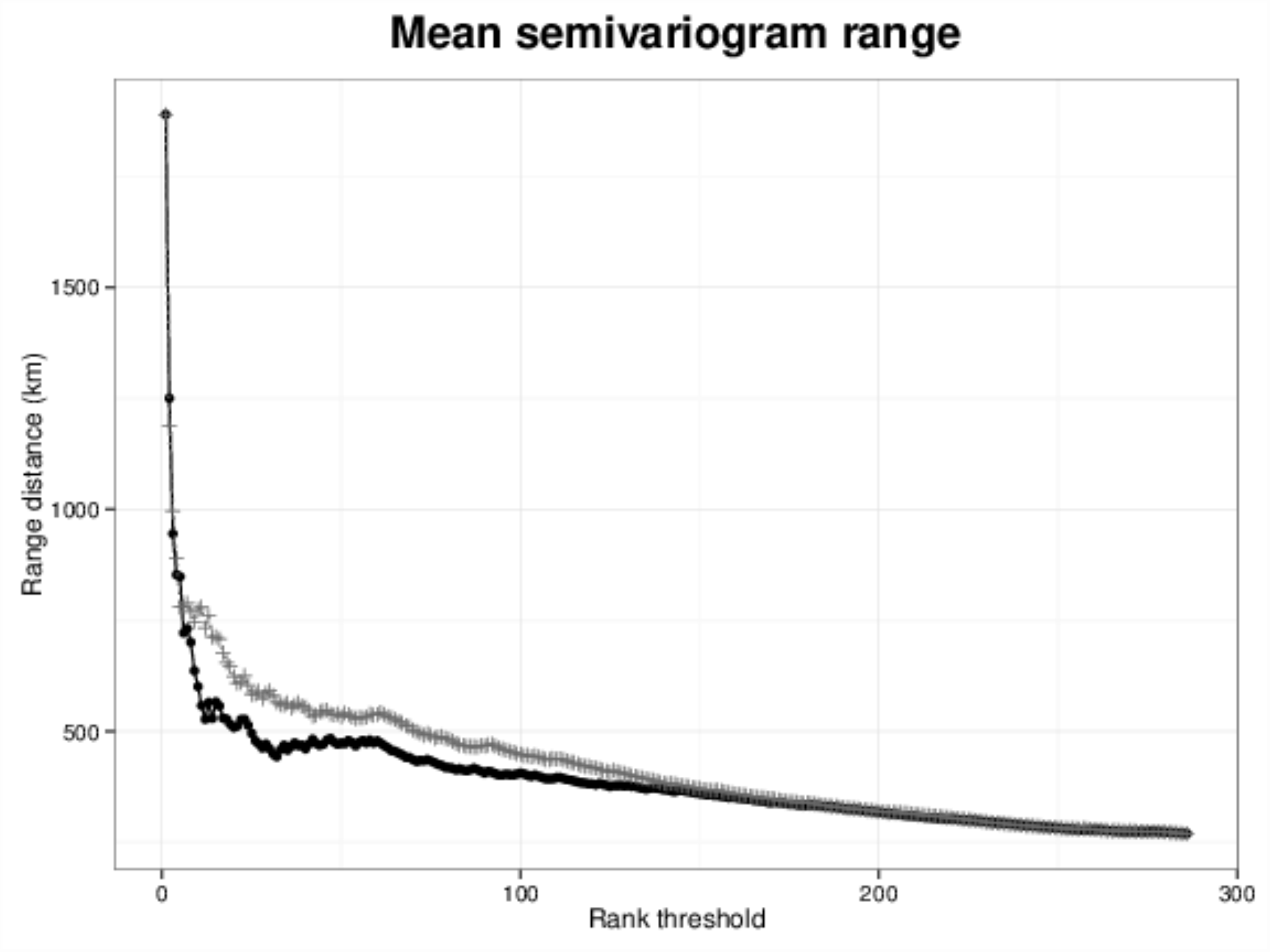}
\includegraphics[width=0.45\textwidth]{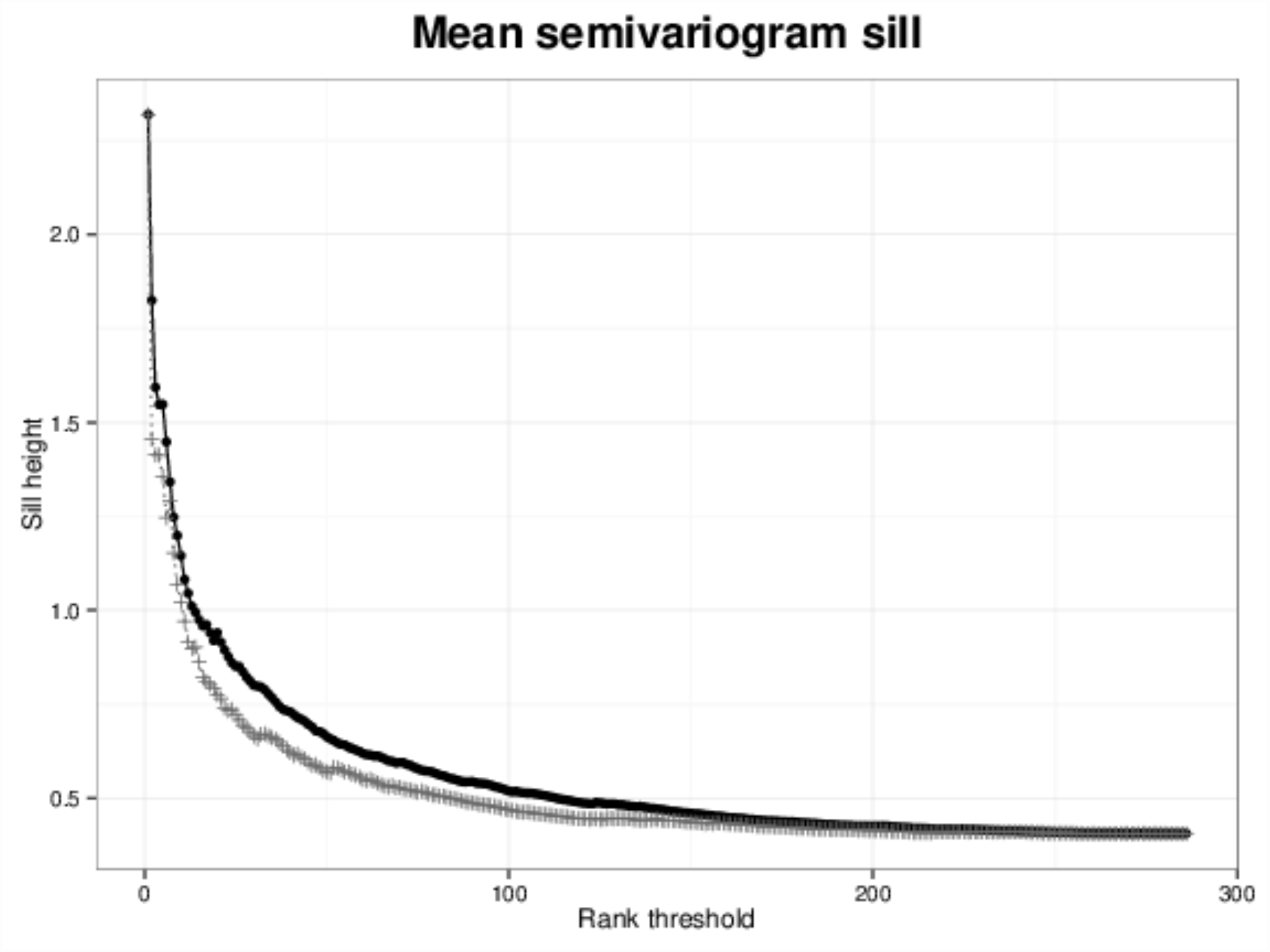}
\includegraphics[width=0.45\textwidth]{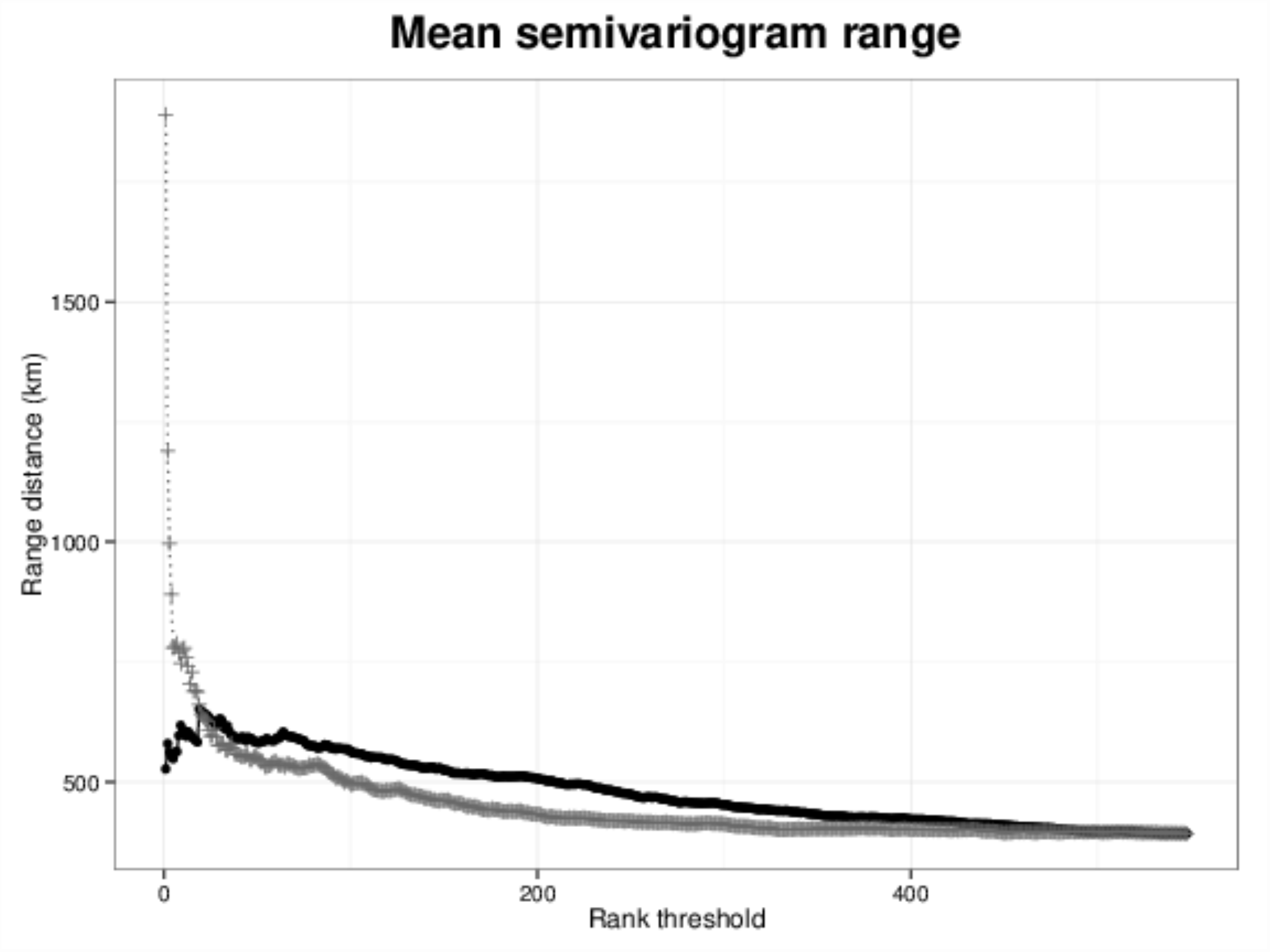}
\includegraphics[width=0.45\textwidth]{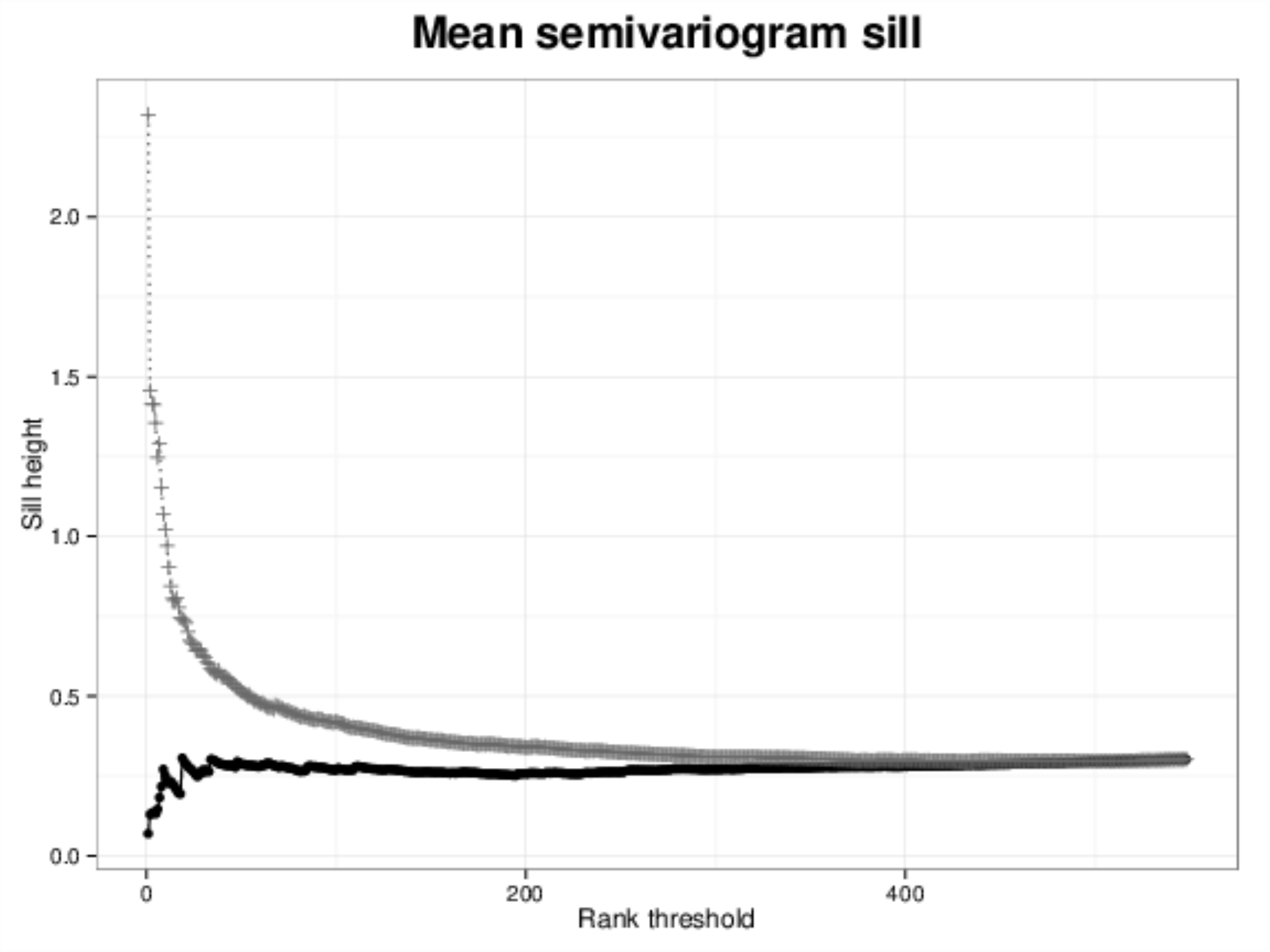}
\caption{ {\bf Average semivariogram properties
for groups of top $N$ ICD-9 codes by the NB2 and Moran's I methods.}
We show the mean range (left) and mean sill (right) for both the
NB2 method (black points; top: t-test implementation, bottom: log odds implementation) 
and Moran's I statistic (grey crosses) plotted
against the rank threshold $N$.
Compared to Moran's I, the NB2 method with t-test implementation
ranks more highly spatial variation with
smaller ranges (autocorrelation within smaller distances) and larger sills 
(greater variance), whereas the NB2 method with log odds implementation
ranks more highly spatial variation with larger ranges (autocorrelation within
larger distances) and smaller sills (lower variance).
}
\label{Fig3}
\end{figure}

We compare the results of semivariogram modeling for the highest ranked ICD-9 codes 
using the two NB2 method implementations 
to the semivariogram models for the highest ranked ICD-9 codes using
Moran's I statistic.  Specifically, we compare average semivariogram model properties 
between groups of the top $N$ ranked ICD-9 codes for
increasing values of $N$ using the two NB2 rankings and Moran's I ranking.  We will refer
to $N$ as the rank threshold.
In the top of Figure \ref{Fig3} (left), we show the mean semivariogram range 
vs the rank threshold $N$ using the NB2 method with t-test comparison (black) and Moran's I statistic (grey).  This shows
the average distance range within which the incidence rates are autocorrelated for the 
top $N$ ranked ICD-9 codes by each method.  
For example, the mean semivariogram model range for the top 25 ICD-9 codes ranked by the NB2 
method is 495 km vs 625 km for the top 25 Moran's I statistic rankings. 
For the top 100 ICD-9 codes, the mean semivariogram model range is 404 km and 509 km for 
the NB2 method with t-test comparison and Moran's I statistic, respectively.  Generally, the NB2 method 
using the t-test comparison implementation ranks
more highly ICD-9 codes showing spatial variation with smaller ranges, 
or smaller areas of autocorrelation.

In the bottom of Figure \ref{Fig3} (left), for comparison we show the same plot
of the highest ranking semivariogram range properties for the NB2 method with
log odds comparison (black) and Moran's I statistic (grey).  Generally, the NB2 method
using the log odds comparison implementation ranks more highly ICD-9 codes showing
spatial variation with larger ranges, or larger regions of autocorrelation.


In the top of Figure \ref{Fig3} (right), we show the mean
semivariogram sill vs the rank threshold $N$ using the NB2 t-test implementation (black) and Moran's I
statistic (grey).  This essentially shows an estimate of the average variance in incidence
rates across the US for the top $N$ ranked ICD-9 codes by each method.  For example,
the mean sill for the top 25 ICD-9 codes ranked by the NB2 
method is 0.85 vs 0.67 for the top 25 Moran's I statistic rankings. 
For the top 100 ICD-9 codes, the mean semivariogram model sill is 0.52 and 0.42 for 
the NB2 method t-test implementation and Moran's I statistic, respectively.  
In this case the NB2 t-test implementation generally ranks more highly ICD-9 codes with larger variance in the 
incidence rates across the US.

In the bottom of Figure \ref{Fig3} (right), we show the same plot of the highest ranking
semivariogram sill properties for the NB2 method with log odds comparison (black)
and Moran's I statistic (grey).  Generally, the NB2 method using the log odds comparison
implementation ranks more highly the autocorrelated ICD-9 codes with smaller variance.

\section*{\uppercase{Discussion}}

There are many possible explanations for spatial patterns in the
incidence rates of ICD-9 EMR data, and the rank ordering of ICD-9 codes with
the described methods does not attempt to attribute any inferred pattern to 
a specific cause or suggest that the spatial variation is due to a physical environmental 
factor.  Rather, we provide here a spatial autocorrelation method that can be
implemented in multiple ways depending on the type of spatial pattern of interest.
This flexibility is useful given that different categories of underlying factors as well as 
categories of disease can manifest as different spatial patterns, as we discuss below.

Incidence levels of diseases are influenced by a variety of factors, including:
\begin{itemize}
\item \textit{Physical environment} --- Some diseases are known 
to be related to the physical environment.
\item \textit{Socioeconomic environment} --- The incidence levels of
some diseases are impacted by socioeconomic or  regional cultural differences.
\item \textit{Structural environment} - The incidence levels of some diseases 
reflect in part geospatial differences in insurance,  provider billing or reimbursement patterns,
local regulations, and related factors.
\end{itemize}

\begin{figure}[!ht] 
\centering


\includegraphics[width=0.475\textwidth]{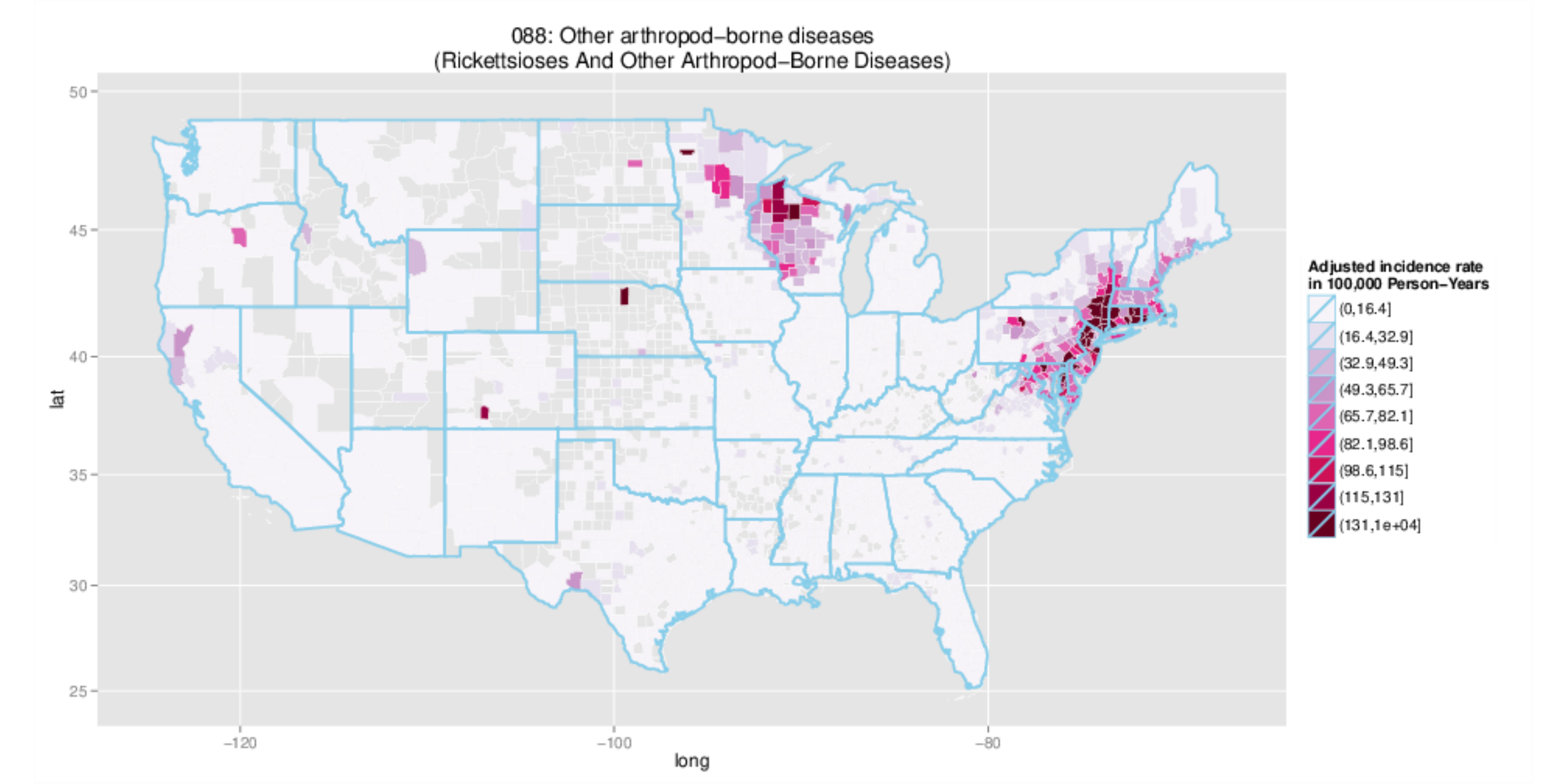}
\includegraphics[width=0.475\textwidth]{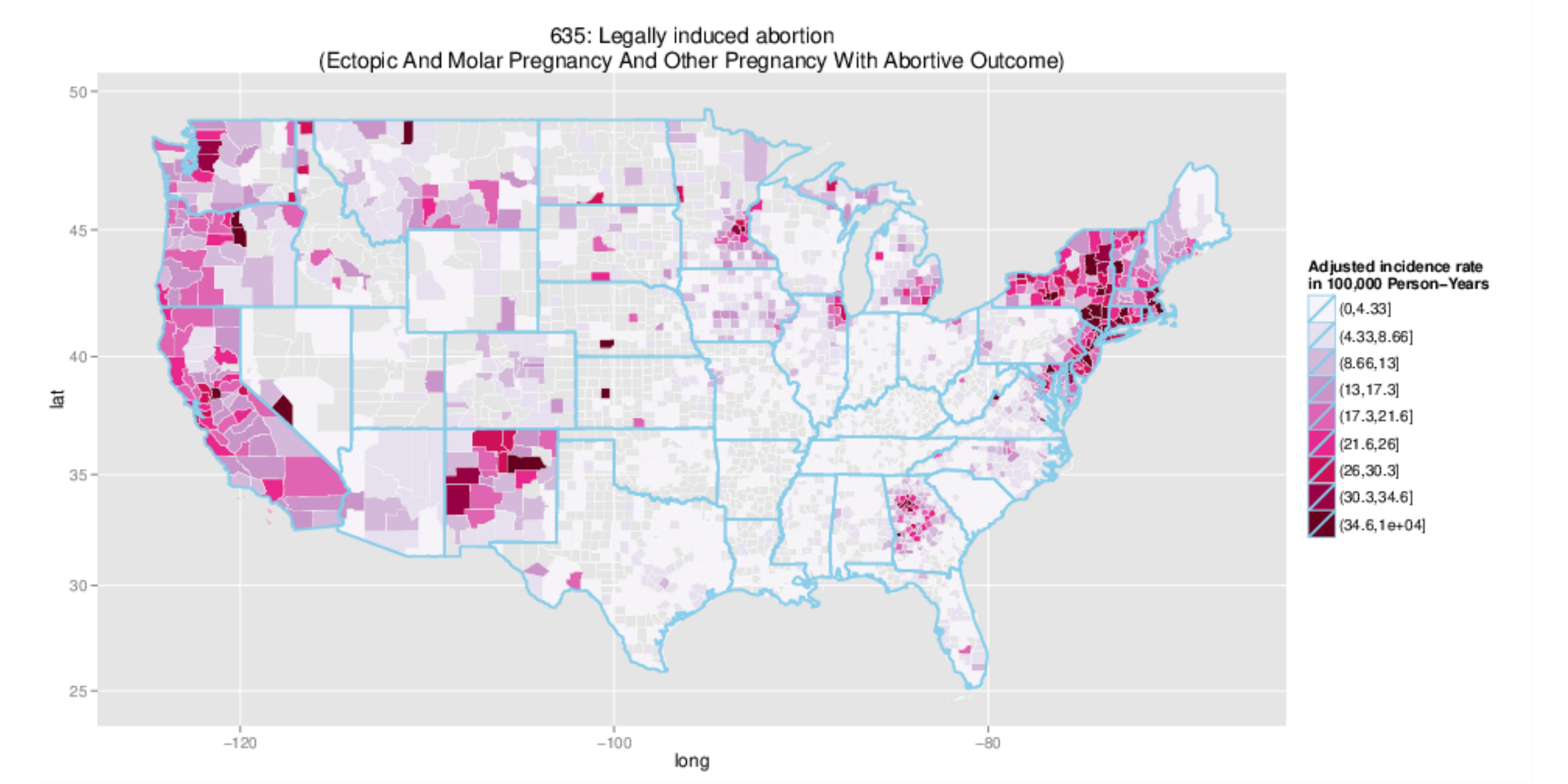}
\includegraphics[width=0.475\textwidth]{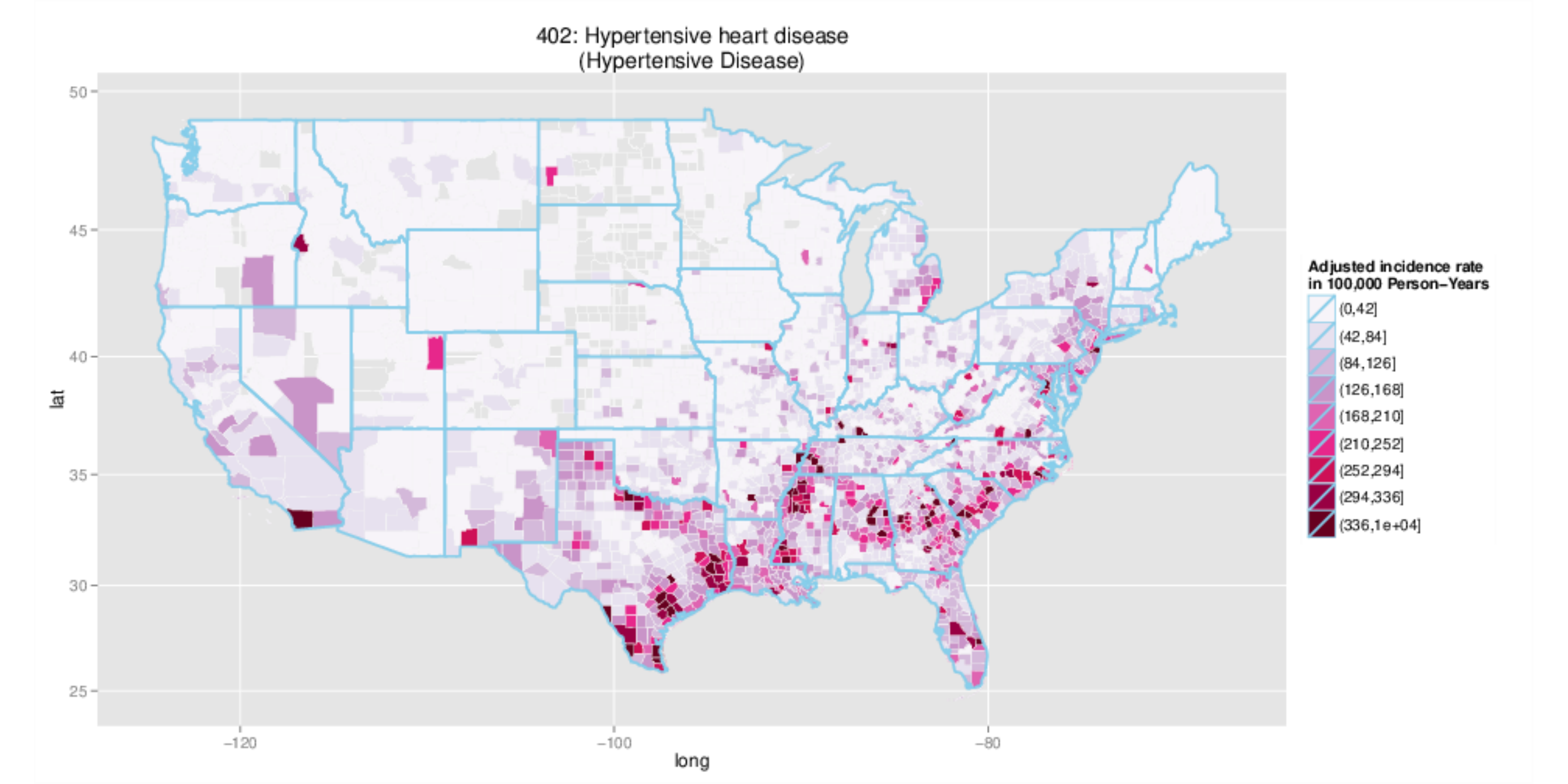}
\includegraphics[width=0.475\textwidth]{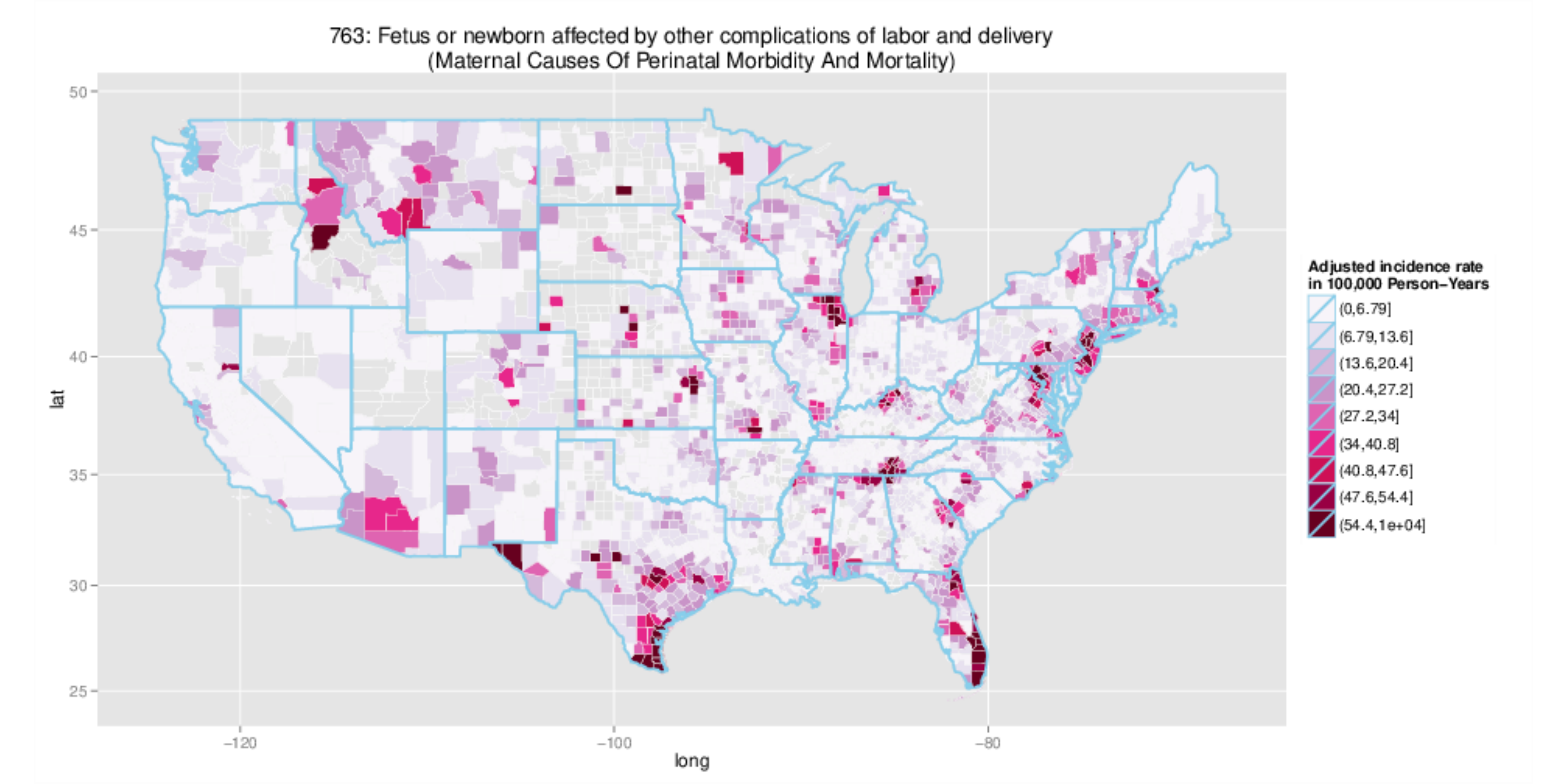}
\caption{{\bf Incidence maps for several ICD-9 codes with different
types of spatial variation.} We show here
088: Other arthropod-borne diseases, (top left);
635: Legally induced abortion, (top right); 
402: Hypertensive heart disease, (bottom left);
and 763: Fetus or newborn affected by other 
complications of labor and delivery, (bottom right).
}
\label{Fig4}
\end{figure}

We show several incidence rate maps in Figure \ref{Fig4} as examples of patterns corresponding to these
three types.  These ICD-9 codes are all ranked in the top 25 according to at least 
one implementation of the NB2 method.  In the top left is a map showing ICD-9 code 088: Other arthropod-borne diseases, 
which includes Lyme disease, a disease carried by ticks and known to have a regional concentration in the northeastern
US and western Wisconsin areas.  
We consider this as an example of an ICD-9 code with spatial variation 
due to the \textit{physical environment}.  In the top right is a map showing ICD-9 code 635: Legally 
induced abortion.  The spatial variation for this ICD-9 code shows clear delineation of the borders 
between states, which is likely to be due to differences in the \textit{structural environment}.  
The delineation is particularly apparent on the borders between California and Nevada
and New York and Pennsylvania.  
In the bottom left of Figure \ref{Fig4} is a map showing ICD-9 code 402: Hypertensive heart disease, 
which is the ICD-9 code
ranked highest by the NB2 method.  The spatial variation shows a pattern of higher
incidence rate across a large crescent in the southern US.  Given this cross-state regionally concentrated
pattern, we define this to be an example of differences in the \textit{socioeconomic environment}.
In the bottom right we show a map of ICD-9 code 763: Fetus or newborn affected by other complications
of labor and delivery, which is not easily classified as the previous
three examples.

\subsection*{Characteristics of disease categories}
\begin{figure}[!ht] 
\centering

\includegraphics[width=0.95\textwidth]{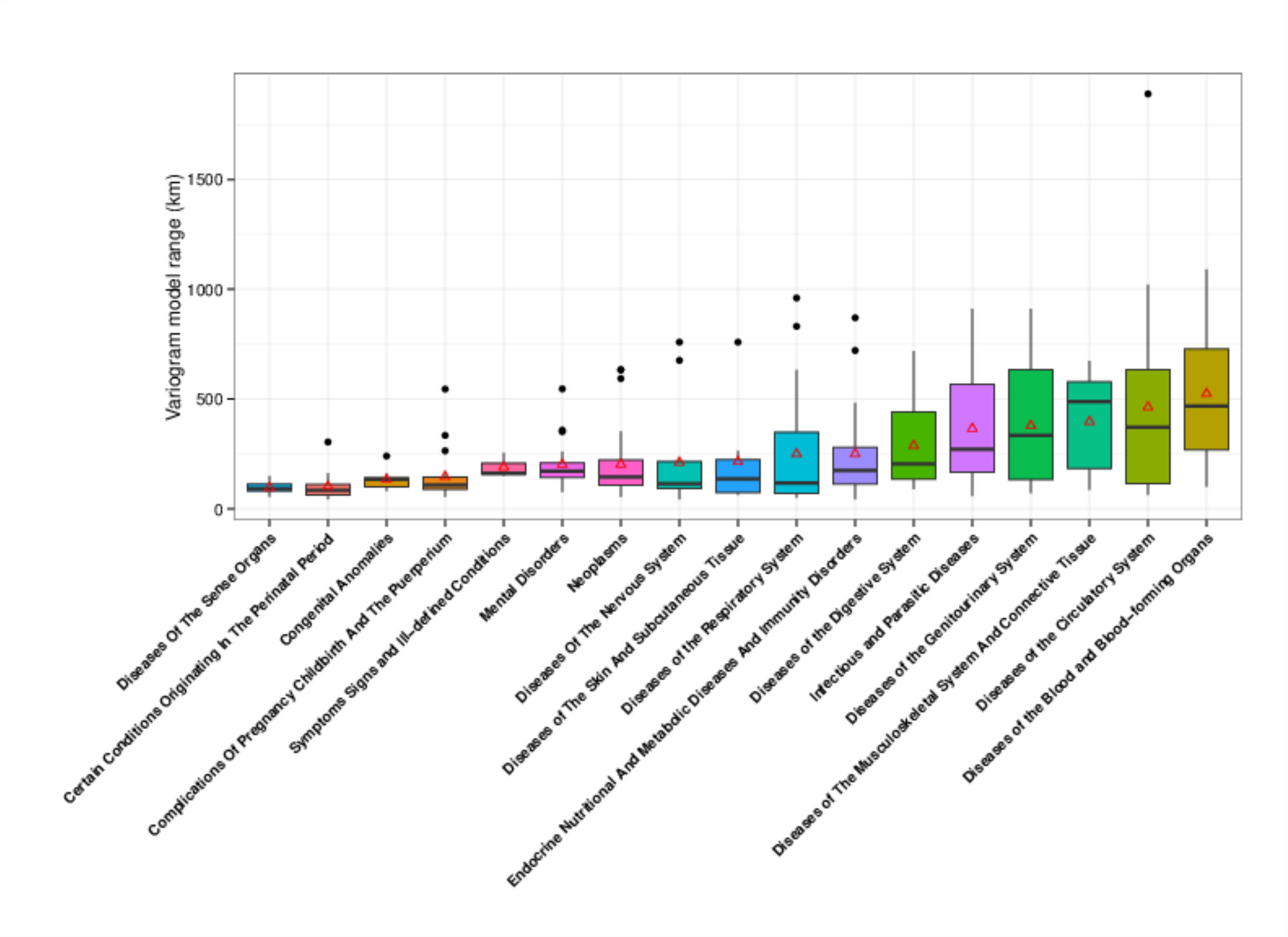}
\caption{ {\bf Box plot of variogram model ranges for categories of ICD-9 codes.}
The categories are in increasing order by the average variogram model range 
(marked by red triangles) for each category.
}
\label{Fig5}
\end{figure}

In building semivariogram models describing the spatial variation for each ICD-9 code,
we also looked at the model properties for categories of disease collectively. 
We grouped the ICD-9 codes according to standard categories, for example, 
001-139 Infectious and Parasitic Diseases, 140-239 Neoplasms, etc.
For each group we found the mean semivariogram model range, excluding
ICD-9 codes where the semivariogram model range fit failed to iterate beyond the initial
starting value, which leaves 286 individual ICD-9 codes in 17 categories.
In Figure \ref{Fig5}, we show a box plot of the semivariogram model
ranges for each category ordered by increasing mean range.  

There is no correlation between the mean semivariogram model ranges of categories and 
the number of ICD-9 codes grouped into each category.
The categories with the fewest remaining ICD-9 codes are Symptoms, Signs, 
and Ill-defined
Conditions (3 codes), Diseases of the Blood and Blood-forming Organs (6 codes), and
Diseases of the Skin and Subcutaneous Tissue (8 codes).  The categories of
Neoplasms and Infectious and Parasitic Diseases have the most codes with 29 and 26 codes,
respectively.  However, the diseases with the smallest ranges generally also have low mean 
incidence rates across the US.  

Given the variation of typical range values across different disease categories, one or the
other presented implementation of the NB2 method may be appropriate for the detection
of a spatial pattern for the type of disease of interest.

\section*{\uppercase{Conclusions}}
We have described here a bootstrap method that can be implemented
multiple ways for detecting patterns in spatial variation based upon a region's neighbors.
The Neighbor-based Bootstrapping (NB2) method is a procedure for quantifying how 
much more accurate an estimate of the value of interest is based on values from 
bootstrapped neighboring units than bootstrapped randomly chosen units.

We have compared two implementations of the NB2 method to Moran's I statistic for measuring
spatial autocorrelation.  Generally, the NB2 method and Moran's I statistic are in 
rough agreement although with some scatter and interesting differences.  Looking at the rank orderings of ICD-9 code
county incidence rates across the US ranked by the NB2 method and by Moran's I statistic
shows that, by choosing one or the other implementation of the NB2 method, we can favor 
spatial variation with autocorrelation within smaller distances or of larger scale.  
Compared with Moran's I statistic,
the NB2 method allows more flexibility in controlling the type of spatial autocorrelation of interest.

Compared to Moran's I statistic, the NB2 method using the t-test comparison ranks more highly the
ICD-9 codes that appear to have multiple small clusters over a region whereas the NB2 method using
a log odds comparison ranks more highly the ICD-9 codes with large regional gradients.
We also compared the spatial properties of categories of disease by looking at the
mean fitted semivariogram properties of each category and found that 
different categories of disease as a whole may have larger or smaller size scales of autocorrelation,
as measured by average semivariogram model ranges.  For example, ICD-9 codes
related to conditions originating in the perinatal period generally have spatial
variation that is autocorrelated within smaller distance ranges than ICD-9 codes
related to diseases of the blood and blood-forming organs.  Given this difference
in spatial variation scale, one or the other implementation of the NB2 method may be more appropriate
depending on the category of disease.

\section*{\uppercase{Acknowledgments}}

This material is based in part upon work supported by the National Science Foundation under Grant Number 1129076 and the The National Cancer Institute (NCI) under Contract  NIH/Leidos Biomedical Research, Inc. 13XS021 / HHSN261200800001E.

This work made use of the Open Science Data Cloud (OSDC), managed by the Open Commons Consortium (OCC) and funded in part by grants from the Gordon and Betty Moore Foundation \cite{Grossman2012}.

\section*{\uppercase{Author Disclosure Statement}}

No competing financial interests exist.



\bibliography{abbrev-refs.bib}

\clearpage
\section*{\uppercase{Supplementary Material}}

In these supplementary materials we describe the results of the NB2 rankings by disease
category and describe and spatial patterning for those ICD-9 codes ranked highly 
by either the NB2 t-test or odds implementation.

\paragraph*{Infectious and Parasitic Diseases (001$-$139)}
Of all the disease categories,
Infectious and Parasitic Diseases have the highest average NB2 rank (most spatially significant) using the paired t-test implementation.
In the box plot of Figure \ref{Fig5} this category includes 26 remaining ICD-9 codes with a mean semivariogram distance range of 365 km.
Four of these diagnoses codes are ranked in the top 10 by the t-test implementation of the NB2 method.  Two of these are animal related diseases, including
088: Other arthropod-borne diseases, which includes tick-borne Lyme disease with a geographic pattern
in the Northeastern US and Wisconsin areas as shown in Figure \ref{Fig4} and 115: Histoplasmosis, 
also known as Ohio Valley disease, which is a fungal infection commonly spread by bat and bird droppings in soil.  
Histoplasmosis is also spatially interesting by the odds implementation of NB2, ranked 34.5.
The other two codes are
sexually transmitted infections which mainly spread the southern regional area of the US:
131: Trichomoniasis and  
099: Other venereal diseases, which also shows a pattern due to the structural environment 
with high incidence in the state of Michigan. 

\paragraph*{Neoplasms (140$-$239)}
Neoplasms as a group on average are ranked in the bottom half of the diseases categories (less spatially significant) by both implementations of NB2.
The highest ranked ICD-9 code is 219: Other benign neoplasm of uterus, ranked 6 by t-test, with a small range of 90 km.
There are 29 codes in Figure \ref{Fig5} with a mean range of 200 km.
The codes with the largest range (outliers in Figure \ref{Fig5}) are 173: Other malignant neoplasm of skin, ranked 130 by t-test and 98 by odds,
with a large range of 635 km, affecting the southeastern US and the western and northeastern coastal areas.  
ICD-9 codes 189: Malignant neoplasm of kidney and other and unspecified urinary organs and 
203: Multiple myeloma and immunoproliferative neoplasms also have large outlying ranges but have low
NB2 ranks.
	
\paragraph*{Endocrine, Nutritional, Metabolic, and Immunity Disorders (240$-$279)}
This category of diseases has the third highest average NB2 rank using the t-test, 9th using odds, a mean range of 250 km, 
and several diagnosis codes in this category are ranked in the top 50 by NB2 with t-test.
These include 268: Vitamin D deficiency, ranked 19, which has clusters of higher incidence in central Virginia, northern New York, southern
Texas, central New Mexico, and the northwestern coast of the US;
259: Other endocrine disorders, ranked 24, which has small clusters in the southern and western US
and also shows a pattern due to the structural environment with high incidence in Michigan; 
257:  Testicular dysfunction, ranked 29, which has a large range (outlier in Figure \ref{Fig5} at 720 km) with broadly affected
regional areas in the south and also the west coast;
and 266: Deficiency of b-complex components, ranked 38, 
has high incidence clusters in the Kentucky, Tennessee, Alabama, Georgia area.
%
Additionally, 251: Other disorders of pancreatic internal secretion, which is ranked 52,
is also an outlier in Figure \ref{Fig5} with a large range of 720 km. 
Using the odds NB2 implementation, 272: Disorders of lipoid metabolism is ranked highest at 11, and
250: Diabetes mellitus is next highest, ranking 22.5.	
	
\paragraph*{Diseases of the Blood and Blood-forming Organs (280$-$289)}
This category of diagnosis codes has the largest average semivariogram range (525 km). 
The top ranked codes by NB2 are 281: Other deficiency anemias, ranked 15 with t-test and 63 with odds, 
280: Iron deficiency anemias, ranked 60 with t-test and 42 with odds, 
and 285: Other and unspecified anemias, ranked 101 with t-test and 21 with odds, 
all of which show large-scale gradients with higher incidence in the south and southeast.

\paragraph*{Mental Disorders (290$-$319)}
The 19 ICD-9 codes in Figure \ref{Fig5} have a smaller range on average (200 km).
The top NB2 ranked codes are 
309: Adjustment reaction, ranked 65 with odds and 87.5 with t-test, 
which is most prominent in the northeast, northwest, and Michigan;
302: Sexual and gender identity disorders, ranked 65 by t-test and 196 by odds, 
with high incidence in the states of Georgia, South Carolina, and Nevada;
and 304: Drug dependence, ranked 66 by t-test and 172.5 by odds, with high incidence in regions of Kentucky, Tennessee, and Alabama.
The highest range outlier is 290: Dementias, 
which shows a strong signature of differences in the structural environment with high incidence in the states
of Massachusetts and California and sharp state borders.

\paragraph*{Diseases of the Nervous System (320$-$359)}
This category of ICD-9 codes has a similar mean distance range to Neoplasms and Diseases of the
Skin and Subcutaneous Tissue (210 km).
The top NB2 ranked codes are 356: Hereditary and idiopathic peripheral neuropathy, ranked 93.5 using the t-test and 92.5 using odds, 
which has a higher than average range (760 km) reflecting a gradient of high incidence in the south and southeast;
and 347: Cataplexy and narcolepsy, ranked 107 using the t-test and 370 using odds, which has clusters of high incidence primarily in central Ohio, 
central Indiana, Michigan, and South Carolina.
The ICD-9 code 357: Inflammatory and toxic neuropathy also has a large outlying range (675 km)
with higher incidence generally in Texas, Nevada, and the southeastern US.

\paragraph*{Diseases of the Sense Organs (360$-$389)}
The ICD-9 codes in this category were ranked relatively low by NB2 (least spatially significant)
and also have the smallest average range of autocorrelation (95 km).
The highest ranking code is 367: Disorders of refraction and accommodation, ranked 35 by the t-test and 89.5 by odds, 
which has higher incidence generally in the northwestern border states, central Illinois, southern
Georgia, and in the states of Connecticut and Nevada.

\paragraph*{Circulatory System (390$-$459)}
This category of ICD-9 codes is the third highest
ranked as spatially interesting using the mean of the NB2 t-test method results.  Many codes in this category are ranked
in the top 20.  This category also has the second largest average semivariogram model range (465 km),
meaning large spatial areas are autocorrelated.
The top ranking codes using the odds NB2 implementation include
401: Essential hypertension, ranked 1 using the odds method and 171 using the t-test method,
which shows a gentle gradient affecting the whole US with higher incidence in the southern region;
424: Other diseases of endocardium, ranked 7 using odds and 21 using the t-test,
which also shows a gradient with higher incidence in the south and also some clustered areas;
and 414: Other forms of chronic ischemic heart disease, ranked 12 by odds and 108, 
which is also more prevalent in the south.
The top ranking codes using the t-test are
402: Hypertensive heart diseases, ranked 1 using the t-test and 19 using odds,
which has high incidence in a large arc across the southern US shown in Figure \ref{Fig4};
413: Angina pectoris, ranked 7 using the t-test and 9 using odds, 
which has high incidence primarily in the south central US, Texas, and Florida;
and 411: Other acute and subacute forms of ischemic heart disease, ranked 13 using the t-test and 30 using odds, 
which also has high incidence in the south central and east north central region US.

\paragraph*{Respiratory System (460$-$519)}
The NB2 method ranks the diagnosis codes in this category on average highly, the second highest group by
the t-test and 4th by odds.
The mean semivariogram range for the 25 codes shown in Figure \ref{Fig5} is 250 km with a large
spread in the distribution.
Several codes are ranked in the top 25:
514: Pulmonary congestion and hypostasis, ranked 18 by t-test and 21 by odds, 
which has higher incidence in the central and southern US;
487: Influenza, ranked 21 by t-test and 8 by odds, 
which has higher incidence in the southern US, excluding Florida, and a large range (830 km, high outlier);
and 476: Chronic laryngitis and laryngotracheitis, ranked 25 by t-test and 231.5 by odds, 
which has smaller (55 km semivariogram range) clusters across the US.
Allergic rhinitis (ICD-9 477) is also ranked second highest using the NB2 odds method, 47th using the  t-test,
and affects all of the US but is generally more prominent in the southern US.
Chronic bronchitis (ICD-9 491) has the highest range (960 km) and is ranked 33 with higher incidence
generally in the central US.

\paragraph*{Digestive System (520$-$579)}
Diseases in this category are mid ranked by NB2.
The top ranked codes using the odds method are
530: Diseases of the esophagus, ranked 5 by odds and 149 by t-test,
which shows generally higher incidence in the south and southeastern US;
535: Gastritis and duodenitis, ranked 13 by odds and 56 by t-test,
which also is more prevalent in the southern areas with some smaller clustering;
and 564 Functional digestive disorders not elsewhere classified, ranked 25.5 by odds and 98 by t-test,
which again shows broad higher incidence in the south and southeast.
The top ranked codes using the t-test method are
520: Disorders of tooth development and eruption, ranked 10.5 by t-test and 32 by odds, 
which shows sharp state outlines indicating differences in the structural environment 
with high incidence in Tennessee,
Wisconsin, Pennsylvania, New Jersey, Delaware, Maine, and Michigan in particular;
and 533: Peptic ulcer site unspecified, ranked 39 by t-test and 210.5 by odds, which generally has higher incidence
in small clusters around the central and southern US . 

\paragraph*{Genitourinary System (580$-$629)}
This category is ranked relatively high by both NB2 implementations (3rd by odds and 5th by t-test) 
and also has a higher average semivariogram range (380 km).
The highest rank ICD-9 code in this category is 
627: Menopausal and postmenopausal disorders, ranked 4 by odds and 70 by t-test,
which has high incidence in large areas, particularly in the southern states as well as Montana and Michigan.
Other ICD-9 codes with high ranks include
616: Inflammatory disease of the cervix, vagina, and vulva, ranked 6 by odds and 75 by t-test,
which shows higher incidence generally in the southern states;
599: Other disorders of urethra and urinary tract, ranked 10 by odds and  261.5 by t-test,
which also shows higher incidence in southern states;
and 602: Other disorders of the prostate, ranked 9 by t-test and 202 by odds,
which has high incidence regions primarily in southern Georgia, Texas, and the areas of Michigan, Ohio,
Pennsylvania, New Jersey, and Maryland.
Redundant prepuce and phimosis (ICD-9 605) is ranked 27 by t-test and 164 by odds and has several clusters of high incidence 
primarily in Texas and southern states, as well as central Illinois, northeastern Indiana, and Michigan.
Inflammatory disease of ovary, fallopian tube, pelvic cellular tissue, and peritoneum (ICD-9 614) is ranked 41 by t-test and 64 by odds
with a large range (910 km).

\paragraph*{Pregnancy, Childbirth, and the Puerperium (630$-$679)}
Complications of Pregnancy, Childbirth, and the Puerperium are mid ranked by the NB2 method,
and this category has a small average range (145 km).
The highest ranked codes are 635: Legally induced abortion, ranked 3 by t-test and 48 by odds,
which shows clear state boundaries reflecting different structural environments
and has a subsequently larger outlying range (335 km);
662: Long labor, ranked 17 by t-test and 254.5 by odds, which has dense clusters of high incidence in
Texas, Florida, Georgia, Michigan, Wisconsin, and Montana with very low incidence elsewhere;
and 645: Late pregnancy, ranked 43.5 by t-test and 146 by odds, which has higher incidence generally in the northern
half of the US.

\paragraph*{Skin and Subcutaneous Tissue (680$-$709)}
This category of diseases is ranked in the bottom half of the categories (least spatially
significant) by the NB2 method and has an average semivariogram range of 215 km.
The top codes in this category are
682: Other cellulitis and abscess, ranked 3 with odds and 81 with t-test,
which has higher incidence in the south central US
and 692: Contact dematitis and other eczema, ranked 14 with odds and 165.5 with t-test,
which has higher incidence in large areas in the mid central US, Michigan, California,
Montana, and eastern and southeastern states.
The ICD-9 code 680: Carbuncle and furuncle, ranked 106 by t-test and 182 by odds, has a large 760 km range
(outlier in Figure \ref{Fig5}).  This code has a pattern of higher incidence generally
in one primary main region centered around Louisiana and the Gulf Coast area
and falling off with increasing distance.  A smaller region in the area of eastern Montana
and North Dakota also shows high incidence.

\paragraph*{Musculoskeletal System and Connective Tissue (710$-$739)}
The diagnosis codes in this category have a large average semivariogram range
and are generally ranked of low spatial importance by NB2.
The highest ranking code is 739: Nonallopathic lesions not elsewhere classified, ranked 8 with the t-test and 28 with odds,
which shows clear state borders with high incidence in the northwestern states, particularly 
North Dakota, and also Iowa and Maine.

\paragraph*{Congenital Anomalies (740$-$759)}
The diagnosis codes in this category have a small average semivariogram range
and are also generally ranked of low spatial importance by the NB2 method.
The top code according to NB2 rankings is 743: Congenital anomalies of eye, ranked 128.5 by the t-test and 366.5 by odds. 
The ICD-9 code 754: Certain congenital musculoskeletal deformities has a higher than average 
range for this category (240 km), with high incidence and sharp borders in the state of Michigan.

\paragraph*{Conditions Originating in the Perinatal Period (760$-$779)}
Two codes for perinatal conditions rank in the top 50 by NB2.
Fetus or newborn affected by other complications of labor and delivery (ICD-9 763) is ranked 14 by the t-test and 252.5 by odds
and has small (range 60 km) clusters of high incidence in a dozen or so regions around 
the US, some crossing state borders, and low incidence elsewhere.
Disorders relating to long gestation and high birthweight (ICD-9 766) is ranked 48 by the t-test and 296 by odds
and shows similar smaller clustering in some areas.

\paragraph*{Symptoms and Nonspecific Abnormal Findings (780$-$799)}
On average these codes are ranked low by the NB2 t-test method but are ranked as the category with 
the most spatial autocorrelation by the Moran's I statistic and by the NB2 odds method, several codes ranked in the top 20.
These include
785: Symptoms involving cardiovascular symptoms, ranked 15 by odds and 245.5 by the t-test,
789: Other symptoms involving abdomen and pelvis, ranked 16 by odds and 401.5 by the t-test,
and 786: Symptoms involving the respiratory system, ranked 17 by odds and 431 by the t-test,
all of which cover most of the US with slightly higher incidence in the southern states and Michigan;
The highest ranking ICD-9 code by the NB2 t-test method is 799: Other ill-defined and unknown causes of 
morbidity and mortality, ranked 93.5 using the t-test and 182 using odds,
which has a high level of incidence in the state of Nevada and also has 
clusters of high incidence in Texas, Florida, New York, California, north central Illinois, and central Missouri.

\end{document}